\documentclass[twocolumn,superscriptaddress,amsmath,amssymb,aps,sort&compress,prx]{revtex4}

\usepackage[utf8]{inputenc}

\usepackage{times,color,amsthm,graphics,graphicx,bm,bbm,dcolumn}
\usepackage{epsfig}
\usepackage{comment}
\usepackage{graphicx}
\usepackage{xcolor}
\usepackage{booktabs}
\usepackage[colorlinks,urlcolor=black,citecolor=blue,linkcolor=blue]{hyperref}
\usepackage{units}
\usepackage{multibbl}

\setcitestyle{square}

\begin{document}

\title{Imprinting persistent currents in tunable fermionic rings}

\author{G.~Del~Pace}
\altaffiliation[Present address: ]{Institute of Physics, EPFL, 1015 Lausanne, Switzerland}
\email[E-mail: ] {delpace@lens.unifi.it}
\affiliation{Istituto Nazionale di Ottica del Consiglio Nazionale delle Ricerche (CNR-INO) and European Laboratory for Nonlinear Spectroscopy (LENS), University of Florence, 50019 Sesto Fiorentino, Italy}

\author{K.~Xhani}
\affiliation{Istituto Nazionale di Ottica del Consiglio Nazionale delle Ricerche (CNR-INO) and European Laboratory for Nonlinear Spectroscopy (LENS), University of Florence, 50019 Sesto Fiorentino, Italy}

\author{A.~Muzi~Falconi}
\affiliation{Department of Physics, University of Trieste, 34127 Trieste, Italy}
\affiliation{Istituto Nazionale di Ottica del Consiglio Nazionale delle Ricerche (CNR-INO) and European Laboratory for Nonlinear Spectroscopy (LENS), University of Florence, 50019 Sesto Fiorentino, Italy}

\author{M.~Fedrizzi}
\affiliation{Istituto Nazionale di Ottica del Consiglio Nazionale delle Ricerche (CNR-INO) and European Laboratory for Nonlinear Spectroscopy (LENS), University of Florence, 50019 Sesto Fiorentino, Italy}

\author{N.~Grani}
\affiliation{Istituto Nazionale di Ottica del Consiglio Nazionale delle Ricerche (CNR-INO) and European Laboratory for Nonlinear Spectroscopy (LENS), University of Florence, 50019 Sesto Fiorentino, Italy}
\affiliation{Department of Physics and Astronomy, University of Florence, 50019 Sesto Fiorentino, Italy}

\author{D.~Hernandez~Rajkov}
\affiliation{Istituto Nazionale di Ottica del Consiglio Nazionale delle Ricerche (CNR-INO) and European Laboratory for Nonlinear Spectroscopy (LENS), University of Florence, 50019 Sesto Fiorentino, Italy}

\author{M.~Inguscio}
\affiliation{Department of Engineering, Campus Bio-Medico University of Rome, 00128 Rome, Italy}
\affiliation{Istituto Nazionale di Ottica del Consiglio Nazionale delle Ricerche (CNR-INO) and European Laboratory for Nonlinear Spectroscopy (LENS), University of Florence, 50019 Sesto Fiorentino, Italy}

\author{F.~Scazza}
\affiliation{Department of Physics, University of Trieste, 34127 Trieste, Italy}
\affiliation{Istituto Nazionale di Ottica del Consiglio Nazionale delle Ricerche (CNR-INO) and European Laboratory for Nonlinear Spectroscopy (LENS), University of Florence, 50019 Sesto Fiorentino, Italy}

\author{W.~J.~Kwon}
\altaffiliation[Present address: ]{Department of Physics, Ulsan National Institute of Science and Technology (UNIST), Ulsan 44919, Republic of Korea}
\affiliation{Istituto Nazionale di Ottica del Consiglio Nazionale delle Ricerche (CNR-INO) and European Laboratory for Nonlinear Spectroscopy (LENS), University of Florence, 50019 Sesto Fiorentino, Italy}

\author{G.~Roati}
\affiliation{Istituto Nazionale di Ottica del Consiglio Nazionale delle Ricerche (CNR-INO) and European Laboratory for Nonlinear Spectroscopy (LENS), University of Florence, 50019 Sesto Fiorentino, Italy}

\begin{abstract}


Persistent currents in annular geometries have played an important role in disclosing the quantum phase coherence of superconductors and mesoscopic electronic systems. Ultracold atomic gases in multiply connected traps also exhibit long-lived supercurrents, and have attracted much interest both for fundamental studies of superfluid dynamics and as prototypes for atomtronic circuits. Here, we report on the realization of supercurrents in homogeneous, tunable fermionic rings. We gain exquisite, rapid control over quantized persistent currents in all regimes of the BCS-BEC crossover through a universal phase-imprinting technique, attaining on-demand circulations $w$ as high as $9$. 
High-fidelity read-out of the superfluid circulation state is achieved by exploiting an interferometric protocol, which also yields local information about the superfluid phase around the ring. In the absence of externally 
introduced perturbations, we find the induced metastable supercurrents to be as long-lived as the atomic sample. Conversely, we trigger and inspect the supercurrent decay by inserting a single small obstacle within the ring. For circulations higher than a critical value, the quantized current is observed to dissipate via the emission of vortices, i.e., quantized phase slips, which we directly image, in good agreement with numerical simulations. The critical circulation at which the superflow becomes unstable is found to depend starkly on the interaction strength, taking its maximum value for the unitary Fermi gas. Our results demonstrate fast and accurate control of quantized collective excitations in a macroscopic quantum system, and establish strongly interacting fermionic superfluids as excellent candidates for atomtronic applications.





\end{abstract}

\maketitle


\section{INTRODUCTION}

In a conducting ring pierced by a magnetic field, the electronic wavefunctions are augmented by a phase winding enforced by the associated vector potential \cite{Byers1961,Bloch1965}. 
The most striking manifestation of such geometric phase factor is the quantization of the magnetic flux enclosed by the loop into discrete values, inherently associated with the emergence of persistent diamagnetic currents around the ring \cite{Deaver1961,Doll1961,Onsager1961,Byers1961,Bloch1965}. 
In $\mu$m-sized metallic rings, such persistent currents are the hallmark of mesoscopic physics, as they are underpinned by the electronic phase coherence throughout the entire system \cite{Bluhm2009,Bleszynski2009}. In superconducting or superfluid loops, characterized by 
macroscopic phase coherence \cite{TinkhamBook}, circulating supercurrents appear whenever particles experience a gauge field -- which in neutral matter may be synthetically created \cite{Dalibard2011,Goldman2014} -- as an outcome of the famous Aharonov-Bohm effect \cite{Aharonov1959} and the continuity of the wavefunction around the loop. 
Supercurrents are extremely long-lived, and are for instance exploited in the operation of superconducting high-field electromagnets with vanishing energy dissipation \cite{Kadin1999book}. 


In neutral superfluids, such as helium or ultracold atomic condensates, metastable supercurrents \cite{Leggett1999,Mueller2002} may also flow in the absence of a gauge field: 
Even though the ground state of a 
superfluid ring has zero current, a supercurrent can be excited, e.g., by externally imposing a dynamical phase gradient around the loop. Since the phase of the superfluid wavefunction must wind around the ring by an integer multiple of $2\pi$, only discrete supercurrent states can exist \cite{Bloch1975}, denoted by an 
integer winding number $w \neq 0$. States with different $w$ are 
associated with $w$-charged vortices 
enclosed in 
the superfluid. In multiply-connected ring geometries, vortices are preferentially trapped inside the ring at vanishing density, and topologically distinct $w$-states are thus separated by energy barriers \cite{Bloch1975}. As a result, excited \emph{persistent} current states in superfluid rings are long-lived and generally insensitive to disorder, thermal fluctuations and even particle losses.


Atomic persistent currents have been realized in weakly interacting Bose-Einstein condensates (BECs) confined in toroidal traps \cite{Ryu2007,Moulder2012,Beattie2013}. 
Studies of the long-time dynamics of persistent currents 
in BECs have revealed their metastable and quantized character \cite{Moulder2012}, identifying quantized phase slips as the excitations connecting states with different winding numbers with one another \cite{Moulder2012,Wright2013_driving,Kumar2017}.
Recently, long-lived single-winding ring currents have been produced for the first time by stirring strongly interacting Fermi gases \cite{Cai2021}, and have been observed to survive even when the system is driven out of the superfluid phase and back.
Yet, many open questions remain about the detailed physical processes governing the persistent current decay, especially in fermionic pair condensates where multiple excitation mechanisms may compete or cooperate with one another. For tightly confined rings, coherent and incoherent phase-slippage processes are thought to dominate the decay in different temperature and interaction regimes \cite{Dubessy2012,Polo2019,Kunimi2019,Mehdi2021}, sharing intriguing analogies with the glassy dynamics and avalanches observed in certain disordered systems \cite{Ferrero2021}.

%
Besides providing an important playground for the investigation of artificial quantum matter, ultracold atomic rings are also a promising platform for quantum technologies.
An annular atomic superfluid embodies the minimal realization of a matter-wave circuit, providing an elementary building block for future atomtronic devices \cite{Amico2021, Amico2021_atomtronic} and neutral current-based qubit implementations \cite{Aghamalyan2015}. Persistent currents in ring condensates could be employed for quantum sensing, offering an atomtronic analogue of superconducting quantum interference devices (SQUIDs) \cite{Ramanathan2011, Ryu2013, Eckel2014_hysteresis, Ryu2020, Kiehn2022} or new-generation inertial force sensors \cite{Gustavson1997,Navez2016,Pandey2019} thanks to the Sagnac effect. For any technological application, a high-degree of control over persistent currents is essential; in particular a rapid deterministic creation and a high-fidelity read-out of the current state at the single quantum level are of utmost importance.

In this work, we realize fermionic superfluids in tunable, homogeneous ring traps in different regimes of the crossover from a molecular BEC to a Bardeen-Cooper-Schrieffer (BCS) superfluid, and control their circulation  at the single quantum level by optically imprinting an adjustable dynamical phase.
By employing an interferometric probe to inquire the winding number of the ring, we
demonstrate our protocol to be effective throughout the crossover in creating on-demand supercurrents of tunable circulation. 
These states are observed to persist as long as the atomic sample, demonstrating circulating currents to be particularly robust against particle losses. 
On the other hand, we promote the persistent current decay by inserting a local defect in the ring.
Whereas we find low-circulation states to be insensitive to the introduced perturbation, for circulations higher than a critical value we observe the supercurrent to decay via the emission of quantized vortices, which 
cause phase slippage 
leading to the reduction of the phase winding around the ring down to a lower stable value. 
The onset of dissipation is triggered when the lowest lying excitations become energetically accessible, exposing the different characters of paired Fermi 
superfluids.



\section{TUNABLE FERMIONIC RINGS}

\begin{figure}[t!]
\centering
\vspace{0pt}
\includegraphics[width=1\columnwidth]{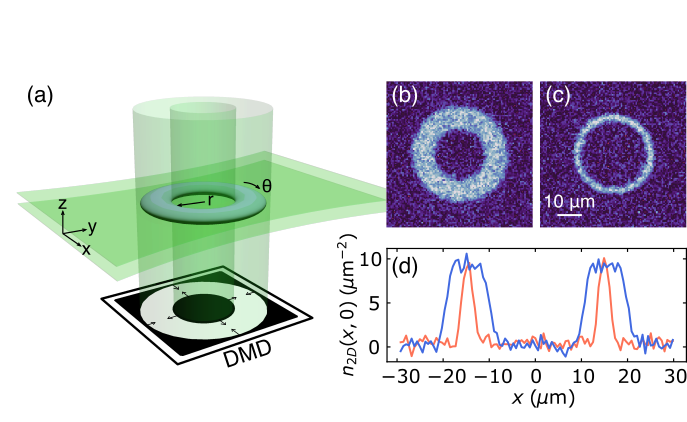}
\caption{Tunable homogeneous rings with ultracold fermionic superfluids. (a) Sketch of the experimental setup: a blue-detuned TEM$_{01}$-like beam propagates horizontally and provides a tight harmonic trap 
in the vertical $z$ direction, whereas a DMD-made repulsive potential sculpts a tunable ring-shaped box-like confinement in the $x-y$ plane. The latter is imaged onto the atomic cloud by a high-resolution microscope objective, omitted in the figure. 
(b)-(c) \textit{In situ} images of a unitary Fermi superfluid confined in two ring geometries with different radial thickness, comprising around $7.5 \times 10 ^3$ and $2.7 \times 10 ^3$ fermion pairs, respectively. The images correspond to a single experimental acquisition of the atomic density. (d) Density cut along the $y=0$ diameter of the rings in (b) and (c) are shown as blue and orange  lines, respectively, obtained by averaging over $20$ realizations.}
\label{Fig1}
\end{figure}

In our experiment, we produce fermionic superfluids of $^{6}$Li atom pairs in 
different regimes across the BEC-BCS crossover by varying the interatomic $s$-wave scattering length $a$ between the two lowest hyperfine states in the vicinity of a Feshbach resonance. 
We focus on three distinct superfluid regimes of a molecular BEC, a unitary Fermi gas (UFG) and a BCS superfluid with coupling strength $1/k_F a \simeq (3.6, 0, -0.4)$, where $k_F = \sqrt{2 m E_F}/\hbar$ is the Fermi wave vector, with $\hbar=h/2\pi$ the reduced Plank constant, $m$ the mass of $^6$Li atom, and $E_F/ h \simeq \{8.2, 9.4, 9.0\}$\,kHz the Fermi energy in the three regimes, respectively.
The superfluid ring is realized by the combination of a tight harmonic confinement along the vertical direction and sculpted optical potentials in the $x-y$ plane [Fig.~\ref{Fig1}(a)].
We adjust the harmonic vertical trap 
frequency $\nu_z$ 
to ensure a similar relative confinement of
$h \nu_z / E_F \sim 0.05$ throughout all the explored interaction regimes, thus keeping our ring superfluid three-dimensional.
The ring-shaped trapping potential in the $x-y$ plane is tailored by a digital micromirror device (DMD) illuminated with blue-detuned light to provide a repulsive optical potential. The DMD provides us the flexibility for arbitrarily tuning the superfluid ring geometry, both in terms of mean radius and thickness [Fig.~\ref{Fig1}(b,c)]. 
In particular, the latter can be reduced down to a Gaussian full width at half maximum (FWHM) of $2.7(1) \, \mu$m in the radial direction [Fig.~\ref{Fig1}(c)], reaching similar conditions to those explored in Ref.~\citenum{Cai2021}. In such geometry, the size of the ring becomes comparable with the typical size $\xi_0$ of fermionic pairs in the strongly interacting regime.
At unitarity 
$\xi_0 \sim 1/k_F \simeq 0.3 \, \mu$m, and $\xi_0$ grows as $\xi_0 \simeq 1/k_F \exp (-\pi / 2 k_F a)$ 
toward the BCS limit, taking a value $\xi_0 \simeq 0.6 \, \mu$m at $1/k_F a = -0.4$. 
Even deeper in the BCS regime, the evolution of the superfluid phase around the ring would become essentially one-dimensional in terms of the azimuthal spatial coordinate, allowing for future explorations of one-dimensional out-of-equilibrium dynamics \cite{ Hofferberth2007,Schweigler2017}. 

Here, we concentrate our attention on the geometry of Fig.~\ref{Fig1}(b), which features internal and external radii of $10$ and $20\,\mu$m, respectively, and comprises $N \simeq 7.5 \times 10^3$ superfluid pairs at a temperature of $T \simeq 0.4 \, T_c$, where $T_c$ is the critical temperature for the superfluid transition in each regime. 
Given the high resolution of the DMD projection setup, the box-trap confinement results in a uniform density both along the azimuthal and the radial direction of the ring [Fig.~\ref{Fig1}(d)], limited by a steep wall with a height much larger than the chemical potential $\mu$ of the superfluid. 

\section{ON-DEMAND EXCITATION AND DETECTION OF SUPERCURRENTS}

\begin{figure}[t!]
\centering
\vspace{2pt}
\includegraphics[width=1\columnwidth]{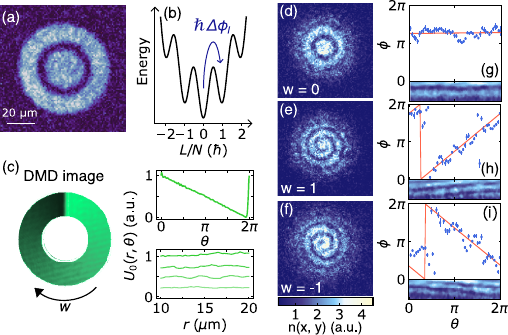}
\caption{Excitation and detection of the circulation state. (a) In situ image of the BEC superfluid in the trap configuration we employ 
to interferometrically probe the ring winding number. (b) Energy landscape of the ring superfluid as a function of the total angular momentum per pair $L/N$. 
The imprinting of an optical phase $\Delta \phi_I$ operates the transition into one of the metastable persistent current states at winding number $w$.
(c) DMD-made light pattern, diagnosed by a secondary imaging camera, employed for the phase imprinting and corresponding light shift $U_0$. 
In the top panel $U_0 (\theta)$ is averaged along the radial direction in the ring, in the bottom one $U_0 (r)$ is averaged over $0.07$ rad across $\theta = 0, \, \pi/2, \, \pi, \, 3 \pi/2$ for decreasing line intensity, respectively. The sharp opposite gradient at $\theta \simeq 2\pi$ has a dimension of $\Delta \theta = 0.15$ rad.
(d-f) Interferograms of the BEC superfluid, obtained from a single absorption image after a TOF expansion of $1.2$ ms without phase imprinting (d) and with a phase imprinting of $\Delta \phi_I \simeq 2 \pi$ in the clockwise (e) and anticlockwise (f) direction. 
(g-i) Fit of the local relative phase $\phi$ extracted from the interferograms. After changing the image into polar coordinates (bottom panel), a sinusoidal fit of each azimuthal slice of the interferogram is performed to extract $\phi$ as the relative phase shift (blue dots on top panels). The polar images comprise the radial region at $1.5 \, \mu$m $<r < 12 \, \mu$m, cut into $100$ azimuthal slices. Symbols (error bars) in the (g-i) top panels represents the weighted average (weighted standard deviation) over the fitted values of $\phi$ and their $1 \sigma $ uncertainty of the fit in bins of $0.125\,$rad size. A linear fit (red solid lines) provides a measurement of the winding number as the slope of the $\phi-\theta$ curve. We obtain $w = 0.01(2)$, $w = 1.00(9)$ and $w = -0.99(9)$ for the (g), (h), (i) panel, respectively.
}
\label{Fig2}
\end{figure}

We excite non-zero circulation states by optically imprinting a dynamical phase on the ring. Optical phase imprinting has been demonstrated to be an effective technique to create solitons \cite{Burger1999, Yefsah2013} and phase imbalance triggering Josephson oscillations \cite{Luick2020}, but has never been employed to inject rotational excitations so far.
We exploit the DMD to create an optical azimuthal gradient with uniform intensity across the radial direction [Fig.~\ref{Fig2}(c)], which we shine onto the superfluid ring for a short adjustable time $t_I$.
When $t_I $ 
is shorter than the characteristic density response time, 
i.e., $t_I < \hbar/\mu$, the light imprints a phase $\phi_I = U_0 \times t_I / \hbar$ onto the 
atomic wavefunctions \cite{Burger1999, Zheng2003}, where $U_0$ is the 
spin-independent potential exerted by the light field on the atomic states. 
The imprinting effectively transfers the superfluid in one of the local minima of the parabolic washboard potential pictorially sketched in Fig.~\ref{Fig2}(b), i.e., into a metastable persistent current state \cite{Moulder2012, Kumar2017}. 
In particular, in the absence of other excitations, when the imprinted phase difference at the gradient discontinuity $\Delta \phi_I$ equals $ 2 \pi w$ a persistent current state with winding number $w$ is excited, characterized by a velocity  $v =  w \hbar / 2mr$ and an angular momentum per superfluid pair $L/N = \,w\hbar$.

To measure the winding number, we carry over 
to fermionic superfluids 
the interferometric technique previously demonstrated for weakly-interacting BECs \cite{Eckel_2014, Corman2014, Mathew2015}.
In analogy to self-heterodyne detection in optics, we exploit a central disk condensate as local oscillator to provide a constant phase reference [Fig.~\ref{Fig2}(a)], and observe the fringe pattern arising after interfering it with the ring during a time-of-flight (TOF) expansion. In the strongly interacting regime, the resulting interference is detected after having tuned $a$ to adiabatically transfer the superfluid into a molecular BEC. 
When the superfluid ring is prepared at rest (see Appendix A for further details), the interferogram displays fringes arranged as concentric rings [Fig.~\ref{Fig2}(d)], revealing an azimuthally uniform phase difference between the two condensates.
On the other hand, when 
a state with finite $w$ is excited in the ring, the interferogram changes into a spiral fringe pattern 
[Fig.~\ref{Fig2}(e,f)], reflecting the linear trend of the wavefunction phase around the ring. In particular, the spiral direction discloses the sign of the winding number $w$, while the number of arms measures its magnitude \cite{Eckel_2014}.
The interferometric detection allows us to quantitatively access the circulation of fermionic superfluid rings, which so far has been probed only by detecting the presence of vortices after a TOF expansion \cite{Cai2021}.
Moreover, the high resolution of the obtained interferograms allows us to extract  
local information on the relative phase $\phi$ between the ring and the disk condensates \cite{Corman2014}. By switching from cartesian to polar coordinates [Fig.~\ref{Fig2}(g-i), bottom] and performing a sinusoidal fit on each azimuthal slice of the interferogram \cite{SM}, we obtain a $\phi-\theta$ curve [Fig.~\ref{Fig2}(g-i), top] whose slope measures the winding number. 
The local imaging of the superfluid-ring azimuthal phase provides an unparalleled resource for future investigations, especially in tightly confined rings. For example, one could investigate the local phase evolution and its fluctuations around the ring when perturbations are added \cite{Eckel_2014, Mathew2015}, when multiple rings are coupled together \cite{Perez2021, Bland2020, Pelegri2019}, or look for phase dislocations which are expected to characterize the interferograms acquired in the strongly interacting 
regime \cite{Pecci2021}.

\section{PERSISTENT CURRENTS ACROSS THE BEC-BCS CROSSOVER}

\begin{figure}[!h]
\centering
\vspace{0pt}
\includegraphics[width=0.91\columnwidth]{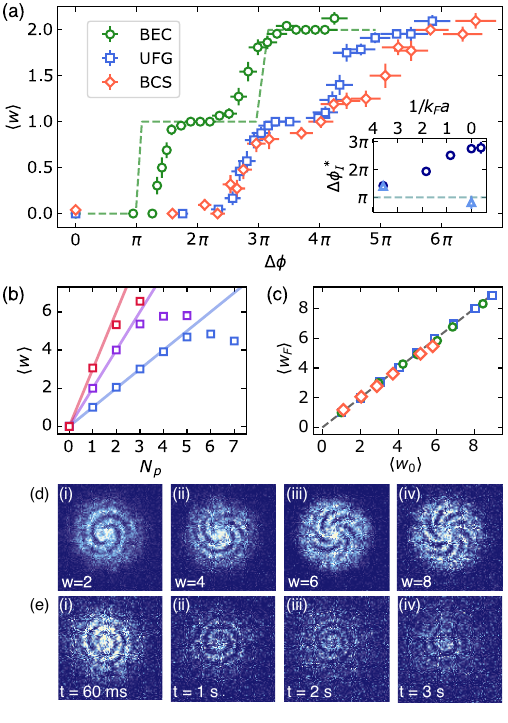}
\caption{Persistent currents with on-demand circulation in the BEC-BCS crossover.
(a) Average winding number $\langle w \rangle $, measured over $20$ interferograms acquired after the imprinting of a phase jump $\Delta \phi_I$ for the three interaction regimes of $1/k_Fa \simeq \{ 3.6, 0, -0.4 \}$, labelled as BEC, UFG and BCS, respectively. 
The interferograms are acquired $60$ ms after the end of the imprinting pulse.
All datasets are obtained by varying the imprinting time at constant $U_0$ in each regime. The green dashed line represents 
GPE numerical simulations of our imprinting protocol, performed under the same experimental conditions of the BEC superfluid. Inset: imprinted phase jump $\Delta \phi_I ^*$ (see text) as a function of the interaction strength for simple gradient imprinting (dark blue circles) and modified by adding a barrier at the discontinuity (light blue triangles). The gray dashed line marks $\Delta \phi_I ^* = \pi$, as expected from GPE numerical simulation. Vertical error bars denote the standard error of mean (SEM), horizontal ones are instead obtained considering the experimental uncertainty of 5\% in the $U_0$
calibration.
(b) Average circulation of a unitary superfluid measured after a variable number $N_p$ of identical imprinting pulses of $w_i = 1$ (blue), $ 2$ (purple), $3$ (red) separated of $10$ ms each from the following. 
The solid lines represent the linear scaling of $\langle w \rangle = w_i \times N_p$.
(c) Average circulation $\langle w_F \rangle $ measured after a long time evolution of the initially imprinted state $\langle w_0 \rangle$ for the three interaction regimes [same notation as in (a)]. 
We measure $\langle w_0 \rangle $ and its SEM $60$\,ms after the imprinting pulse 
and $\langle w_F \rangle $ after $1.5$\,s for $\langle w_0 \rangle \leq 3$, after $1$\,s for $3< \langle w_0 \rangle \leq 6$ and after $0.5$\,s for $\langle w_0 \rangle > 6$ in the BEC and UFG regimes. For BCS superfluids, $\langle w_F \rangle $ is measured after $1$\,s and $0.5$\,s for $\langle w_0 \rangle \leq 3$ and $\langle w_0 \rangle > 3$, respectively.
(d) Interferograms of high circulation states, 
acquired for BCS (i, ii) and UFG (iii, iv) superfluids $60$\,ms after the last imprinting pulse. 
(e) Interferograms of the unitary superfluid in the $w=1$ state as a function of time after the imprinting pulse. Images in (d) [(e)] consist of the average of two [single] 
independent experimental acquisitions.
}
\label{Fig3}
\end{figure}

By imprinting the superfluid phase and reading out the winding number via the interferograms, we gain accurate control over 
the circulation state of the 
rings, which we tune by acting on the imprinting parameters. 
In Fig.~\ref{Fig3}(a) we report the measured mean winding number $\langle w \rangle$ averaged over several experimental realizations of the same imprinting procedure, in the three interaction regimes, as a function of $\Delta \phi_I$. In all superfluids, $\langle w \rangle $ displays a step-like trend, consistent with previous observations with bosonic ring superfluids \cite{Wright2013_driving, Moulder2012} and with
Gross-Pitaevskii equation (GPE) numerical simulations at zero temperature of our imprinting protocol (dashed line) \cite{SM}. Both numerical and experimental BEC data show that $\Delta \phi_I \simeq  2\pi w$ is needed to deterministically excite the circulation state of winding number $w$, but also that a lower imprinted phase is enough to populate it (see Appendix B for further details). 
The switching between $w=0$ and $w=1$ is observed to occur for a slightly larger imprinted phase in the BEC experimental data with respect to GPE results. This is likely due to the collective sound-like excitations which unavoidably affect the superfluid as a consequence of the (optically imperfect) imprinting pulse, 
and 
show a more pronounced effect in experiments with respect to simulations, possibly due to finite temperature. 
In fact, we observe that immediately after an imprinting pulse of any duration the density of the superfluid is depleted at the location of the optical gradient discontinuity (see Appendic C for further details).
We ascribe such a density depletion to a steep, yet not infinitely sharp, intensity gradient in the opposite direction [Fig.~\ref{Fig2}(c)], due to the finite resolution $\sim  1 \, \mu$m of our imaging system. 
Even though the presence of such antigradient has been postulated to completely prevent 
phase imprinting 
to populate persistent current states at finite circulations \cite{Zheng2003}, its particularly small 
spatial extension in our case 
leads to density excitations which decay over a time scale of a few ms into sound waves or vortices. 
The vortices nucleated after the imprinting are observed to survive on top of the macroscopic current for a few hundreds of ms, without perturbing the generated persistent current state, consistent with previous observations in stirred bosonic superfluids \cite{Wright2013_driving}.

In the strongly interacting regime, even though the step-like trend of Fig.~\ref{Fig3}(a) is preserved with a comparable size of the plateaus at non-zero $\langle w \rangle$, we observe that a larger $\Delta \phi_I$ is required to excite the $w = 1$ state. 
We quantitatively estimate such a shift by fitting the first step of the $\langle w \rangle-\Delta \phi_I$ curves with a sigmoidal function to extract the imprinted phase jump $\Delta \phi_I ^*$ necessary to populate the $w=1$ state with $50$\% probability. As reported in the inset of Fig.~\ref{Fig3}(a) in dark blue circles, $\Delta \phi_I ^*$ is observed to monotonically increase with decreasing $1/ k_F a$.
By monitoring the short time dynamics after the imprinting of $\Delta \phi_I = 2 \pi$ in the strongly interacting superfluid at $1/k_F a \simeq 0.9$, we observe that the 
azimuthal phase winding is actually imprinted, confirming the effectiveness of the protocol, but quickly decays out in a few ms (see Appendix C for further details). An imprinted 
$\Delta \phi_I = 2 \pi$ is thus not able to populate the $w=1$ state, and the imprinted angular momentum is dissipated into other excitations, probably 
those triggered by the sharp 
antigradient, which might have a more dramatic effect as we tune the interactions towards the BCS regime, as observed for the imprinting of solitons \cite{Sacha2014}.
To confirm this, we test the imprinting protocol in the BEC and UFG superfluids after modifying the DMD image by inserting a repulsive optical barrier in correspondence of the gradient discontinuity,
as suggested in Ref. \citenum{Perrin2018}, to circumvent the destructive effect of the antigradient on the imprinted current.
In this case, we observe the $\langle w \rangle- \Delta \phi_I$ curves in the two interaction regimes to provide a $\Delta \phi_I^*$ almost comparable with the one expected by GPE simulation (light blue triangles in Fig.~\ref{Fig3}(a), inset). Removing the sharp light gradient allows to drastically reduce the unwanted density excitation produced by the imprinting, yet it leads to a degraded fidelity of only $90\%$ for the inizialization of the $w=1$ state
, likely resulting from the interaction between the moving superfluid and the barrier during the imprinting time \cite{Perrin2018}. For this reason, we choose the simple 
gradient imprinting procedure, which is best suited to deterministically excite non-zero circulation states even in the strongly interacting regime. 
Although previous theoretical and numerical studies pointed out the incapability of the phase imprinting of a simple azimuthal gradient in exciting finite winding numbers in atomic annular superfluids \cite{Zheng2003, Perrin2018}, our results demonstrate it as an effective and versatile method to access on-demand $w$ states. For this, a high-resolution optical pattern is necessary to limit the pollution introduced by the sharp gradient in the opposite direction. 


Through the phase imprinting protocol, we populate the metastable state at $w=1$ 
with a reproducibility $> 99 \%$ in both BEC and unitary regimes, whereas 
we obtain a few-percent lower fidelity in the BCS case.
Higher-$w$ states can then be excited by increasing the imprinting pulse duration. However, in all superfluid regimes, we observe that for increasing $t_I$ the probability to populate a well defined $w$ decreases, while the measured average winding number saturates at about $\langle w \rangle  \simeq 6$, in agreement with GPE numerical simulations \cite{SM}. 
For increasing imprinting time, the more dramatic density excitations 
which follow the imprinting favor the 
nucleation of additional vortices from the innermost ring boundary, removing circulation quanta from the current and setting thus a limit for the highest $w$ which we can populate.
To access even higher $w$ states, we 
keep $t_I \lesssim 500 \, \mu$s and rather increase the number of imprinting pulses. 
Figure~\ref{Fig3}(b) demonstrates the scalability of our imprinting protocol, i.e., the addition of consecutive pulses of a given imprinting time (different colors) produces a linear increment on the measured $\langle w \rangle$.
Even though the multiple phase imprinting protocol becomes ineffective for more than 4-5 pulses, 
this method still surpasses the single-pulse one allowing us to access up to $w_{\mbox{\tiny{max}}} = \{8, 9, 6 \}$ in the BEC, UFG and BCS regime, respectively 
[Fig.~\ref{Fig3}(c,d)].
The whole imprinting procedure takes up to $30$\,ms when we employ $3$ pulses to access $w \geq 5$, demonstrating the multiple 
pulse scheme as a route to excite high circulation states much more rapidly than with stirring protocols \cite{Wright2013_driving}, as employed in Ref.~\citenum{Cai2021} to excite persistent currents in fermionic superfluids.
By monitoring the time evolution of the interferograms in all superfluid regimes 
at long time after the last imprinting pulse, we verify that the imprinted circulation produces a persistent current of constant mean winding number $\langle w \rangle$, which is observed to be as long-lived as the superfluid sample. 
In particular, the $w=1$ current is observed to persist up to $3$\,s with $\langle w \rangle >99 \%$ at unitarity, unaffected by the progressive decrease in atom number [Fig.~\ref{Fig3}(e)].
Circulations higher than $w_{\mbox{\tiny{max}}}$ are also accessible via phase imprinting, but undergo a decay towards states at smaller $w$ in few tens of ms timescale \cite{SM}. 
The maximum stable persistent current $w_{\mbox{\tiny{max}}}$ accessible under our experimental conditions corresponds to a superfluid velocity at the inner ring radius of $4.2$, $4.8$ and $3.2$ mm/s for $w=8$, $w=9$ and $w=6$ in the BEC, UFG and BCS regime, respectively, to be compared with the peak sound speed, estimated as $c_s \simeq \{5.8, 14.1, 14.4\}$ mm/s, respectively \cite{SM}.
In all regimes, the superfluid velocity at $w_{\mbox{\tiny{max}}}$ is observed to be lower than $c_s$, likely due to the excitations introduced by each imprinting pulse, that effectively set an upper bound for the winding number of the persistent current even for the multiple pulse procedure.
In this framework, the relatively lower $w_{\tiny{\mbox{max}}}$ observed in the BCS superfluid is consistent with
pair-breaking excitations lowering the critical velocity for vortices to enter the superfluid ring, effectively reducing the current winding number. As a reference, the critical velocity associated to pair breaking takes a value of $\simeq 9$ mm/s at $1/k_F a = -0.4$ \cite{SM}.

\section{STABILITY OF SUPERCURRENTS AGAINST A POINT-LIKE DEFECT}\label{Sec:Stability}

\begin{figure*}[t!]
\centering
\vspace{0pt}
\includegraphics[width=1\textwidth]{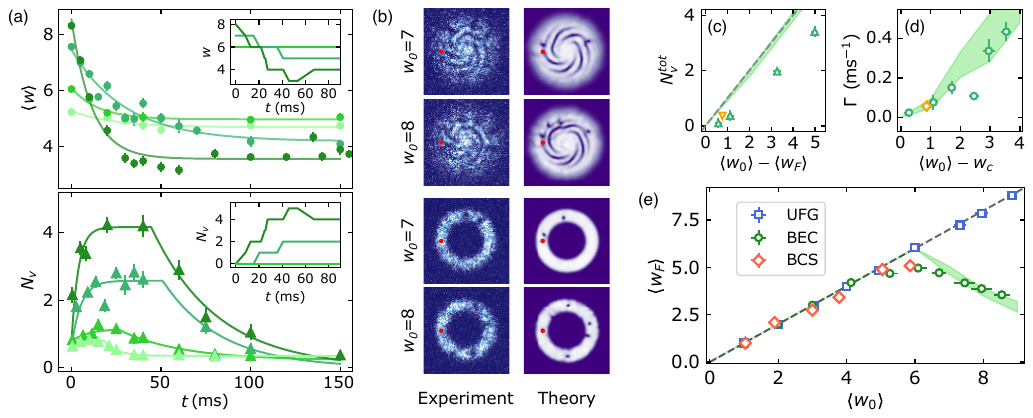}
\caption{Persistent current stability in the presence of a local perturbation. 
(a) Average circulation $\langle w \rangle$ and number of vortices $N_v$ measured after a time $t$ from the introduction of a controlled point-defect in the ring BEC superfluid ($1/k_F a \simeq 3.6$). Symbols with the same color share equal starting experimental conditions, darker shades of green correspond to higher initial circulation. Data in the bottom panel are acquired by ramping up the optical defect $300$\,ms after the excitation of the persistent current, to have on average less than one vortex generated by the imprinting remaining in the ring. Solid lines represent an exponential fit for $\langle w \rangle$, and an exponential charging and discharging fit for $N_v$. Insets: GPE numerical simulations results of the same quantities, plotted using the same color convention.  
(b) Comparison between experimental and GPE simulated interferograms (top) and simple ring TOF (bottom) in the presence of the defect. The position of the defect is marked by a red dot in each image. Experimental interferograms are obtained at $t=20$\,ms, the numerical ones at $t= 40$ ms and $t = 28$\,ms for $w_0 = 7$ and $w_0 = 8$, respectively. The experimental (numerical) images of vortices in the ring are obtained for $t = 5$\,ms and $t = 40$\,ms ($t = 24$\,ms and $t= 53$\,ms) for $w_0 = 7$ and $w_0 = 8$, respectively. The time shift between experimental and numerical images of vortices accounts for the $13$\,ms time interval taken to remove the obstacle in the experiment.
(c-d) Fitted values of the total number of emitted vortices $N_v^{\mbox{\tiny{tot}}}$ and winding number decay rate $\Gamma = A/\tau$, where $A$ is the amplitude and $\tau$ the decay time from the exponential fit, for BEC (green symbols) and $1/k_F a \simeq -0.4$ BCS (orange symbols) superfluids. 
The gray dashed line in (c) marks $N_v^{\mbox{\tiny{tot}}} = \langle w_0 \rangle - \langle w_F \rangle$.
(e) Stability of the circulation for fermionic superfluids (see legend) in the presence of a local defect of height $V_0 \simeq 0.1 \, E_F$ in a BEC and $V_0 \simeq 0.2 \, E_F$ in unitary (UFG) at $1/k_F a \simeq 0$ and BCS superfluids. $\langle w_0 \rangle > w_c$, the final winding number $\langle w_F \rangle$ is obtained from the exponential fit of the current decay, whereas for $\langle w_0 \rangle \leq w_c$ it represents simply the measured average circulation at $t\geq 150$\,ms. In (c-e) panels the results obtained from an exponential fit of the GPE numerical simulations are reported as green shaded regions.
Error bars in panel (a) denote the SEM, while in panels (c-e) they represent the $1 \sigma $ uncertainty of the fit.
}
\label{Fig4}
\end{figure*}

To further investigate the 
connection between persistent currents and vortices arising from their common topological nature, we foster vortex nucleation by 
introducing a controlled optical defect along the ring. 
After having excited a desired circulation state $w_0$, we introduce 
an approximately round obstacle with FWHM$\,= 1.6 \, \mu$m
and height $V_0 \approx 0.1 \, E_F$ for the BEC superfluid, positioned in the midst of the inner and outer edges of the ring trap (see Appendix A for further details). The small defect size, comparable with the characteristic length of the superfluids, is not sufficient to cut the superflow in the radial direction, but rather acts as a local perturbation to the current. By monitoring the time evolution of interferograms as a function of the holding time $t$ in the presence of the defect, we track the average winding number evolution, as reported in Fig.~\ref{Fig4}(a) for the BEC superfluid. In particular, we observe that the defect has no effect on the current up to a critical winding number $w_c = 5$, 
whereas it induces a current decay for $w > w_c$. In the latter case, the interferograms directly display 
the presence of vortices in the superflow, as reported in Fig.~\ref{Fig4}(b). 
We quantitatively study the defect-induced emission of vortices by 
monitoring the TOF expansion of the ring superfluid only [Fig.~\ref{Fig4}(b)], 
after having ramped down the obstacle, and counting the number of vortices $N_v$ as a function of the holding time [Fig.~\ref{Fig4}(a)]. 
After a fast growth, $N_v$ saturates to a value that increases with $w_0$ 
and then decays again as vortices travel outside from the superfluid density. The whole dissipative dynamics develops in a timescale of a hundred of ms, leaving as a result a lower, yet stable current of circulation $w < w_0$ and no vortices. 

To gain more insight into the defect-induced vortex emission dynamics, we perform GPE numerical simulations of the current decay, 
monitoring the time evolution of 
the interferograms [Fig.~\ref{Fig4}(b), left column].
Similarly to the experimental results, numerical simulations involve a critical circulation $w_c = 6$ for the stability of the current [Fig.~\ref{Fig4}(a)], 
above which vortices enter into the superfluid bulk causing the global current to decay [Fig.~\ref{Fig4}(b)]. 
For $w>w_c$, the numerical density profiles of the superfluid ring reveal that 
the defect favors the emergence of a low-density channel which link it to the inner edge of the ring (see Appendix B for further details). Here, the local velocity of the superfluid is observed to increase until it exceeds the local speed of sound and
fosters the nucleation of vortices into the bulk \cite{Xhani2020}.
Each vortex removes one circulation quantum from the ring current, i.e., it induces the phase around the ring to slip by $2\pi$, provoking the current decay as soon as it enters the superfluid density [Fig.~\ref{Fig4}(a), inset in bottom panel]. We note that, contrarily to the experimental data, the detected number of vortices in the numerical interferograms does not decay at long times $t$. Such a discrepancy is likely due to the finite temperature of the experimental system, which is expected to accelerate the vortex escape from the ring density, especially during the interaction with the defect, after completing one turn of the ring.

We perform exponential fits of the decay of $\langle w \rangle $ and the growth of $N_v$, 
parametrizing the observed dynamics in terms of
the circulation decay rate $\Gamma$, 
the final average circulation $\langle w_F \rangle$ and the total number of emitted vortices $N_v^{\mbox{\tiny{tot}}}$ [Fig.~\ref{Fig4}(c-e)]. 
Consistently with a current decay given by vortex nucleation, the total number of emitted vortices is observed to increase with the total drop of the winding number, $\langle w_0 \rangle - \langle w_F \rangle$. Because of technical limitations of the experimental procedure to detect vortices, the measured $N_v^{\mbox{\tiny{tot}}}$ is systematically lower than $\langle w_0 \rangle - \langle w_F \rangle$, whereas numerical simulations show a one-to-one correspondence between the two quantities (shaded region). 
Furthermore, both in experiments and GPE numerical simulations, higher circulations are observed to decay faster and to lower $\langle w_F \rangle$. 
In particular, $\Gamma$ shows a monotonically increasing trend as a function of $\langle w_0 \rangle -  w_c $, reminiscent of the linear scaling of the vortex shedding frequency with the velocity of a moving obstacle in a superfluid at rest \cite{Reeves2015,Kwon2015,Park2018}.
However, a comparison with vortex shedding phenomena should be done carefully, as the superflow velocity in our case is not constant 
over time, because of vortices entering the ring density, and it is higher on the side of the defect facing the inner edge of the ring.
Such an asymmetry plays an essential role in the microscopic route to vortex nucleation, as under our experimental conditions a single vortex per time enters the superfluid bulk, rather than a vortex dipole as for shedding protocols \cite{Kwon2015,Park2018}.

We investigate the effect of the local perturbation on the current in the strongly interacting superfluids as well, changing its height to $V_0/E_F \simeq 0.2$, corresponding to the lowest value of $V_0$ experimentally accessible in this regime (see Appendix A). 
Figure \ref{Fig4}(e) shows the comparison between the defect-induced current decay in the different superfluid regimes.
Similarly to the BEC superfluid, the BCS one presents a critical circulation of $w_c = 5$, whereas we observe no decay for any of the accessible currents in the UFG superfluid. 
Also in the BCS superfluid the current decay is observed to proceed along with the entrance of vortices in the ring density, which happens over a timescale similar to that of the BEC regime [orange symbol in Fig.~\ref{Fig4}(c,d)], suggesting a similar decay dynamics. However, given the higher sound speed $c_s$ in the BCS superfluid, 
the current decay is observed to happen for a smaller relative velocity at the inner ring radius, namely corresponding to $0.18 \, c_s$, instead of $0.46 \, c_s$ as in the BEC. This gap is hardly explained by the different $V_0$ used in the two regimes, as we numerically verified that $w_c$ in BEC superfluids only slightly changes upon increasing $V_0 > 0.1 \, E_F$.
Furthermore, for the same defect height, in the BCS regime $w_c$ is observed to be much smaller than for the UFG superfluid, despite the relatively small difference in $1/k_F a$ and expected $c_s$.
Such observations 
are consistent with 
pair-breaking excitations lowering the critical velocity for vortex nucleation, as suggested also by the lower $w_{\mbox{\tiny{max}}}$ measured in the BCS superfluid and by previous 
measurements of the vortex shedding critical velocity in fermionic superfluids \cite{Park2018}.
On the other hand, the UFG superfluid shows the highest critical circulation in absolute value, even for a relatively higher obstacle with respect to the BEC superfluid, consistent with the expected higher chemical potential and critical velocity 
at unitarity.

\section{CONCLUSIONS AND OUTLOOK}

In this work we have realized tunable fermionic superfluid rings and gained a high degree of control over their persistent current states. We have demonstrated a fast protocol to populate on-demand winding numbers via optically imprinting the superfluid phase, which -- contrarily to other phase-imprinting techniques based on the angular momentum transfer from Laguerre-Gauss beams via Raman transitions \cite{Moulder2012, Ramanathan2011} -- is suitable for application with any atomic species. Furthermore, we applied an interferometric probe to inquire the winding number of the ring fermionic superfluids, which we found to provide reliable and local information on the phase around the ring. These capabilities may be exploited in the future to 
explore 
more complicated geometries, such as double rings tunnel-coupled through a thin barrier to investigate the circulation tunneling \cite{Giamarchi2016,Bland2020,Perez2021,Chestnov2021}, or even ring lattices 
\cite{Pelegri2019}. In particular, a coplanar tunnel-coupled double-ring system would realize the prototype of an atomtronic switch, implementing the atomic analog of a Mooji-Harmans qubit \cite{Bland2020, Mooij2005}.

Furthermore, our work provides the first observation of persistent currents with high winding number in strongly interacting Fermi superfluids, and the first study of their decay induced by a localized obstacle. In the clean system, we observe the current 
be robust against particle losses and as long-lived as the atomic sample. By studying the higher winding number accessible under our experimental conditions, and the critical current in the presence of the defect, we expose the intimate connection between persistent currents and vortices, namely superfluid excitations united by the same topological character. 
In both cases, we observe the supercurrent breakdown to ensue at different critical values across the BEC-BCS crossover, as a consequence of the different dominant excitations – sound or pair-breaking – in different regimes.
A direct extension of our work would be to probe the stability of the current in the presence of multiple defects and eventually in disordered potentials.
Moreover, unitary superfluid currents have shown to be particularly robust against defects, suggesting strongly interacting Fermi superfluids as good candidates for applications in quantum sensing, such as 
atomic gyroscopes, where interactions are expected to enhance their sensitivity \cite{Ragole2016}. 



\begin{acknowledgements}
We thank L.~Amico, T.~Giamarchi, A.~Minguzzi, L.~Pezzè, and the Quantum Gases group at LENS for fruitful discussions. 
This work was supported by the European Research Council under GA no.~307032, the EPSRC under grant no.~EP/R005192/1,
the Italian Ministry of University and Research under the PRIN2017 project CEnTraL, and European Union’s Horizon 2020 research and innovation programme under the Qombs project FET Flagship on Quantum Technologies GA no.~820419 and Marie Sk\l{}odowska-Curie GA no.~843303.
\end{acknowledgements}

\section*{APPENDIX A: EXPERIMENTAL METHODS}
\subsection{Superfluid ring preparation}
\begin{figure}[t]
\centering
\includegraphics[width=1\columnwidth]{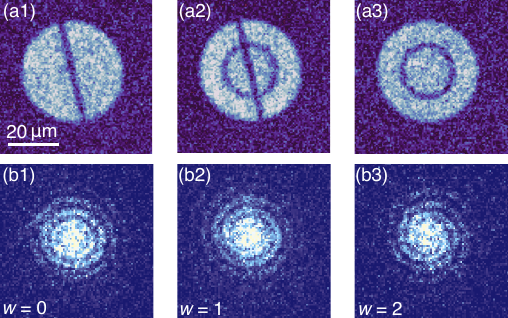}
\caption{(a1-a3) Experimental procedure to create superfluid rings in the $w=0$ state.
The whole trap preparation takes around $130$ ms. Images are obtained as individual \textit{in situ} absorption images of a unitary Fermi gas at different time during the trap loading sequence.
(b1-b3) Interferograms for the ring geometry in Fig.~\ref{Fig1}(c) 
for different optically imprinted winding numbers $w=0,1,2$.
Each panel represents a a single absorption image of the experimental sample.}
\label{Fig: A1}
\end{figure}

To create superfluid rings at rest, we start by preparing a homogeneous circular trap as in Ref.~\citenum{Kwon2021}, complemented by 
a straight barrier across its diameter to hinder the spontaneous excitation of circulating currents in the subsequently created ring 
[Fig.~\ref{Fig: A1}(a1)]. 
We then ramp up the circular barrier which creates the configuration shown in Fig.~\ref{Fig2}(a), used for the interfeometric probing of the current states; once this has reached the final height [Fig.~\ref{Fig: A1}(a2)], we ramp down the straight barrier [Fig.~\ref{Fig: A1}(a3)].
Such procedure allows us to trap fermionic superfluids at rest in rings of different widths and radii, which we can tune by suitably changing the dimensions of the repulsive potentials. 
For the thinnest ring geometries [Fig.~\ref{Fig1}(c)], we initially load a wider ring and subsequently squeeze it by enlarging the inner radius. This procedure allows to trap a large number of atoms in the ring, yielding a clear signal in the interferograms used to determine the supercurrent winding number [Fig.~\ref{Fig: A1}(b1-3)]. In particular, we demonstrate that in the thinnest ring geometry we can deterministically excite the $w=1$ and $w=2$ states [Fig.~\ref{Fig: A1}(b2,b3)], which are observed to persist for several hundreds of ms.

\begin{figure*}[t]
\centering
\vspace{0pt}
\includegraphics[width=2\columnwidth]{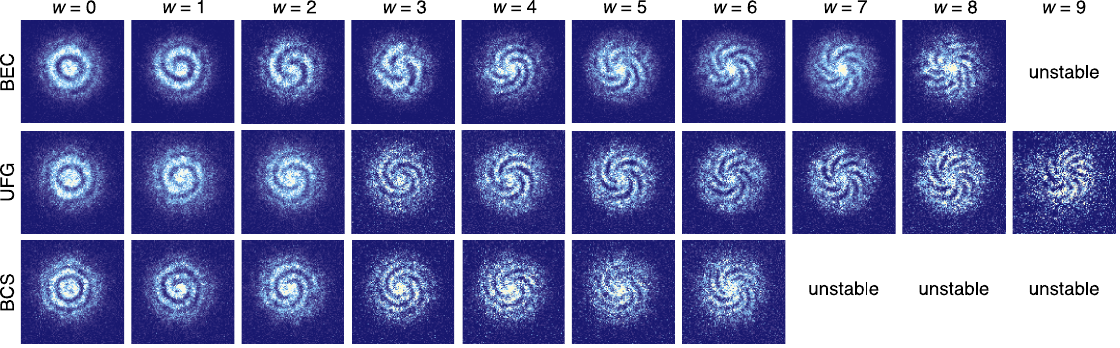}
\caption{Interferograms across the BEC-BCS crossover, 
acquired after a $1.2$ ms TOF expansion. 
Images are obtained by averaging $2-4$ single experimental shots. The colorscales in different rows are differently normalized to compensate for the lower interference contrast in UFG and BCS superfluids.
}
\label{Fig: Gallery}
\end{figure*}

We prepare ring superfluids in the different interaction regimes following the same procedure described in Ref.~\citenum{Kwon2021}. The thermodynamic properties of the superfluids throughout the BEC-BCS crossover are calculated by employing an analytical calculation in the hybrid trap, i.e., harmonic along the $z$ direction and homogeneous in the $x-y$plane, based on the polytropic expansion and reported as Supplemental Material \cite{SM}. We note that a residual magnetic harmonic trap is present in the $x-y$ plane, but its low trapping frequency of $2.5$\,Hz produces negligible effect on the superfluid density. 
The Fermi energy 
is given by $E_F = 2 \hbar \sqrt{\hbar \omega_z N/m (R_{\mbox{\tiny{out}}}^2-R_{\mbox{\tiny{in}}}^2)}$, where $\omega_z \simeq 2 \pi \times \{396, \, 523, \, 529\} $ Hz is the vertical trap frequency in the BEC, UFG and BCS regime, respectively, and $N = 7.5 \times 10^3$ is the number of atoms in the ring of internal and external radii of $R_{\mbox{\tiny{in}}} = 10 \, \mu$m and $R_{\mbox{\tiny{out}}} = 20 \, \mu$m. 
According to the same formulation, the pair chemical potential takes the values of $\mu/h = \{ 1.0, 8.9, 10.0 \}$ kHz, in the BEC, unitary and BCS regime, respectively \cite{SM}. 
These energy scales set the characteristic time and length to $\hbar / \mu \simeq 160 \, \mu$s, $\xi = \hbar / \sqrt{2 m_P \mu} \simeq 0.64 \, \mu$m in the BEC regime, with $m_P = 2m$ the mass of a pair, $\hbar/ \mu \simeq 18 \, \mu$s, and $1/k_F \simeq 0.30 \, \mu$m at unitarity, and $\hbar/ \mu \simeq 16 \, \mu$s and $1/k_F \simeq 0.31 \, \mu$m in the BCS regime. We estimate the pair-breaking velocity as $v_{\tiny{\mbox{pb}}} = [(\sqrt{\Delta^2+\mu^2}-\mu )/m]^{1/2} $ \cite{Comberscot2006, Weimer2015}, where $\Delta = 8/e^2 E_F \exp(\pi/2 k_F a)$ is the pairing gap \cite{ketterle2008}.
In the UFG and BCS regimes, we acquire interferograms after having adiabatically swept the scattering length towards the BEC regime in $50$ ms. We verified that such sweep does not affect the ring current by checking that without any imprinting the interferograms always measure $w=0$. For probing the $w>8$ state at unitarity, we employ instead a fast $3.8\, $ms ramp of the magnetic field to avoid the current to decay during the sweep to BEC, where $w_{\tiny{\mbox{max}}}=8$ \cite{SM}.
A gallery of interferograms at different $w$ in the three interaction regimes is reported in Fig.~\ref{Fig: Gallery}.


\vspace*{-5pt}
\subsection{Light patterns creation}
All the repulsive optical potentials, which we employ for the creation of the ring trap, the realization of the phase imprinting and the introduction of the local defect in the current flow, are obtained by shaping the intensity profile of a 532\,nm blue-detuned laser beam with a DMD.
These optical potentials are 
focused on the atomic cloud through the high resolution ($\sim 1 \,\mu$m) imaging system described in Ref.~\citenum{Kwon2020}. The DMD allows also for a dynamical control of the potential, which we employ in the sample preparation described above, for the phase imprinting and for ramping up and down the obstacle.
To create the 
gradient light profile of 
Fig.~\ref{Fig2}(c) we employ the feedback protocol described in Ref.~\citenum{Kwon2020}. The feedback effectively compensates for the Gaussian profile of the laser beam, producing a radially homogeneous and azimuthally linear light profile.
The sharp gradient in the opposite direction originates instead from the finite resolution of our imaging system, and the feedback procedure cannot correct it.
Each imprinting pulse is realized by turning on the gradient profile in the DMD image for a variable time $t_I \geq 50 \, \mu$s, limit set by the framerate of the device.
In particular, to populate the $w=1$ state we employ $t_I \simeq (130, 150, 170)\,\mu$s for BEC, UFG and BCS superfluids, respectively. We cannot exclude that the working condition of $t_I > \hbar / \mu$
could contribute to the more dramatic effect of the imprinting pulse in the strongly interacting regime, that produces the shift in the $\langle w \rangle - \Delta \phi_I$ curve of Fig.~\ref{Fig3}. We verified at unitarity that increasing the overall gradient height to reduce the imprinting time needed to populate the $w=1$ state down to $ \simeq 110\, \mu$s does not affect the $\langle w \rangle - \Delta \phi_I$ curve shift. These conditions correspond to the lowest $t_I$, i.e., the highest gradient height, accessible by our experimental constraints.
We note that, for a given $t_I$, each imprinting pulse adds the ring superfluid wavefunction of exactly the same phase profile, which results in identical spiral pattern in the interferogram. Thanks to the high reproducibility of the experimental sequence, interferograms obtained under the same condition can be averaged together to increase the fringe contrast.

The obstacle employed in Sec.~\ref{Sec:Stability} is realized by a $4\times 4$ cluster of DMD mirrors, gradually turned on to ramp up its intensity (see Ref.~\citenum{SM} for details on the defect characterization).
In particular, the ramp-up sequence is started $30$ ms after the last imprinting pulse and lasts $13$ ms.
Once the obstacle has reached the final height, it is hold until the end of the experimental cycle for the BEC superfluid, while for the UFG and BCS ones it is ramped down during the sweep towards the BEC regime to acquire the interferograms.
The fact that both the obstacle and the ring trap are realized via the DMD imposes a constraint to the obstacle height in the strongly interacting regime. In particular, $V_0/E_F = 0.2$ corresponds to the lowest laser power impinging on the DMD to be able to trap $7.5 \times 10^3$ atoms in the ring in the BCS regime.
\\
 

\vspace*{-10pt}
\subsection{Vortex detection}

We quantitatively track the vortex nucleation induced by the presence of the defect described in Sec.~\ref{Sec:Stability} by imaging the superfluid density after TOF expansion of the ring only.
For this purpose, we directly load the superfluid in a simple ring trap, added by a repulsive barrier in $\theta = 0$ to prepare it in $w=0$. We excite the desired $w_0$ by employing the same imprinting scheme as for the winding number decay measurement and wait $300$ ms to make sure the vortices created by the imprinting pulse have travelled out of the ring density. We then ramp-up the defect with the procedure already described, hold it for a variable time $t$ and finally ramp it down in $13$ ms to acquire the TOF image. We verified that the average winding number decay is the same whether the obstacle is ramped up after $30$ ms or $300$ ms. 
To maximize the vortex visibility in the ring, we abruptly switch off the vertical trap and we turn off the in-plane ring confinement with a $1$ ms linear rap. We then let the gas to freely expand for other $0.5$ ms before absorption imaging. 
To image vortices in the BCS superfluids we use the same procedure described in Ref.~\citenum{Kwon2021}, employing the trap release already mentioned. 

We note that the measured $N_v^{\mbox{\tiny{tot}}}$ is systematically lower than $\langle w_0 \rangle - \langle w_F \rangle$ (see Fig.~\ref{Fig4}(a,c)).
We mainly ascribe such discrepancy to the technical limitations of the experimental procedure to detect vortices: the $13$ ms removal of the defect is likely to perturb the vortex dynamics, especially in the high $N_v$ regime \cite{Xhani2020}. 
Furthermore, we cannot exclude that the different boundary conditions at the inner ring radius of the geometries with and without the inner disk could affect the vortex emission. 
Finally, we want to stress that, once emitted, vortices leave the superfluid ring density much faster than the one introduced by the imprinting pulse, which are observed to survive on top of the current for several hundreds of ms. It is likely that the presence of the defect could accelerate the vortex escape after the completion of one entire loop, especially at the finite temperature of the experimental superfluid.

\vspace*{-5pt}
\section*{APPENDIX B: THEORETICAL METHODS}

\subsection{Numerical simulation of the phase imprinting}

We perform simulations in the molecular BEC limit 
by numerically solving the time-dependent mean-field 3D Gross-Pitaevskii Equation (GPE) at $T=0$ \cite{SM}. We employ an harmonic 3D potential with a tight confinement along $z$ direction and a hard-wall potential in the $x-y$ plane to create the homogeneous ring trap, using the experimental parameters. 
The ground state wavefunction is found by solving the GPE in imaginary time. To simulate the phase imprinting, we multiply the initial wavefunction by the phase factor $\exp(-i\Delta \phi _I(\theta))$, with $\Delta \phi _I (\theta)=U(\theta)\, t_I/ \hbar$.
In order to model the imprinting potential,  $U(\theta)$  is chosen to have the same azimuthal profile as used in the experiment [Fig.~\ref{Fig2}(c)].
Throughout our studies, we keep the peak value of the imprinting potential 
fixed and equal to the experimental value of $7.8\, \mu$, where $\mu=1.06$ kHz is the numerical chemical potential \cite{SM}, and we vary the imprinting time to access different winding number $w$. 
We find that we a time $t_I=127.4 \, \mu$s is necessary to imprint a total phase of $\Delta \phi_I=2\pi$, i.e., to excite $w=1$, which is consistent with the experimental results in the BEC regime.
\begin{figure}[t!]
\centering
\vspace{11pt}
\includegraphics[width=.9\columnwidth]{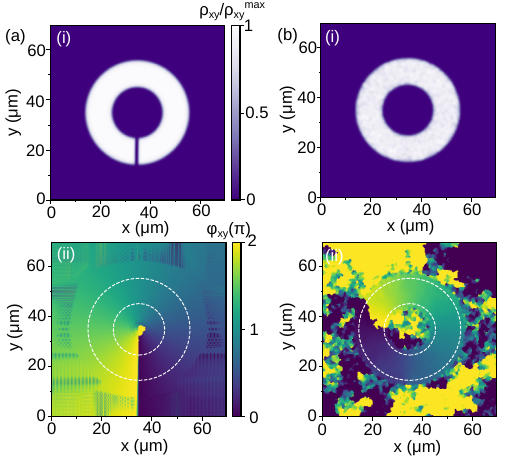}
\caption{The superfluid density [(i)], normalized to its maximum value, and its phase [(ii)] in the $x,y$ plane at $z=0$: (a) immediately after the imprinting of $\Delta \phi_I = 2 \pi$, and (b) after $1$\,s evolution time. The density depletion in (a-i) is due to the finite width of the imprinting potential step. 
The dashed white circles in the phase profiles indicate the ring edges, where the superfluid density takes its bulk value.
}
\label{Fig:w_1_den_phase}
\end{figure}
In Fig.~\ref{Fig:w_1_den_phase}, we show the corresponding $x-y$ plane condensate density $\rho_{xy}$ and phase $\phi_{xy}$ at equilibrium [panel (a)] and $1$ s after the imprinting [panel (b)], extracted from the Madelung representation of the wavefunction $\psi(r,\theta)=\sqrt{\rho(r,\theta)} \exp(i\phi(r,\theta))$.
Consistently with the experimental observations, the condensate density presents a depletion immediately after the imprinting, but it quickly decays in a few ms, leaving only some density fluctuations in the condensate density after $1$ s evolution. A detailed study of the excitations introduced by the imprinting is reported as Supplemental Materials \cite{SM}.
The 2D phase profile clearly shows the presence of a $2\pi$ jump at both times, indicating that a persistent current of $w=1$ is imprinted. We note that initially the phase jump overlaps exactly with the density cut, while at $t=1$\,s it is located at a different angular position because of the superfluid rotation. 

\begin{figure}[t!]
\centering
\vspace{11pt}
\includegraphics[width=.98\columnwidth]{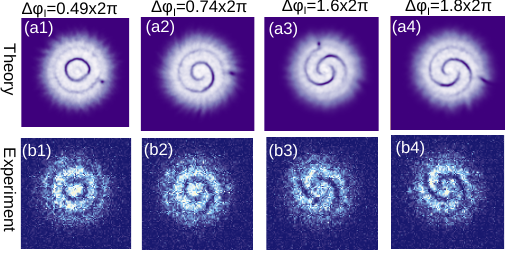}
\caption{Comparison between numerical (a) and experimental (b) BEC interferograms for different imprinted phases $\Delta \phi_I$ (see upper labels), corresponding to the imprinting times: [(a1, b1)] $t_I=64 \, \mu $s, [(a2, b2)] $t_I=96 \, \mu $s, [(a3, b3)] $t_I=204 \,  \mu $s and [(a4, b4)] $t_I=230\,  \mu $s. Both numerical and experimental interferograms are acquired after an evolution of $60$\,ms.} 
\label{Fig:theo_exp_spiral}
\end{figure}

In order to directly compare the numerical and the experimental data, we implement the same experimental interferometric technique. 
We obtain numerical interferograms by quickly removing the ring boundaries in less than $1$ ms and letting the ring and disc condensates to interfere. 
For increasing imprinting time, and thus imprinted phase, we observe the numerical interferograms to change as shown in Fig.~\ref{Fig:theo_exp_spiral}, where they are compared with experimental interferograms acquired for the BEC superfluid under the same imprinting conditions.
When we imprint a non-zero phase $\Delta \phi_I < \pi$, 
the interferograms show almost circularly concentric rings [(a1)], which characterizes the $w_0=0$ 
state. 
In this case, the numerical interferogram reveals the presence of a vortex, 
which carries the imprinted angular momentum. By  further increasing $\Delta \phi _I$ we observe that the $w=1$ or $w=2$ persistent currents can be excited even for $\Delta \phi _I<2\pi$ [(a2)-(b2)] or $\Delta \phi _I<2 \times 2\pi$ [(a3)-(b3),(a4)-(b4)], respectively, and vortices or antivortices resulting from the decay of the initial density depletion are also involved to balance the imprinted angular momentum. The simulated interferograms match well the experimentally recorded ones, and clearly show the presence of vortices for imprinted phases $\Delta \phi _I\neq 2\pi w$, with $|w|=0,1,2$. 
Such vortices are also visible in experiments, but their unambiguous identification on top of interference fringes is difficult.

\subsection{GPE simulation of the defect-induced current decay}

We extend the 3D GPE numerical study at $T=0$ to simulate the defect-induced current decay under the experimental condition of Sec. \ref{Sec:Stability}.
The studies in this section are performed by imprinting an ideal phase gradient to our initial condensate state, 
in order to avoid any density excitations from the imprinting procedure. 
In the absence of the defect the current produced with such imprinting method is observed to be stable for all the investigated $w_0\leq 10$.
\begin{figure*}[t!]
\centering
\vspace{11pt}
\includegraphics[width=.98\textwidth]{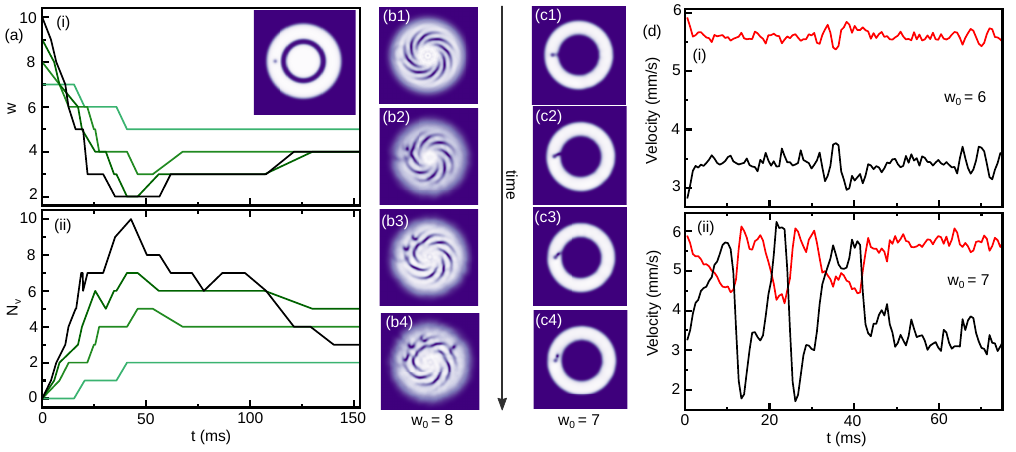}
\caption{Numerical study of the current decay in the presence of a local defect.
(a) Time evolution of the winding number (i) and of the number of emitted vortices 
for different circulation states in the range $6< w_{0}\leq 10$. By employing the confining geometry shown as inset, we compute the numerical interferograms to extract $w$ and $N_v$. Here, the same color convesion of Fig.~\ref{Fig4} is used.
(b1)-(b4) numerical interferograms obtained at different time during the vortex emission process for an imprinted circulation $w_{\mathrm{0}}=8$. The spiral arms gives information about the winding number $w$, the dots indicate instead the vortices.  
The interferograms are extracted at $t=0.3$ ms (b1: $w = 8$, $N_v = 0$), $t=8.9$ ms (b2: $w = 7$, $N_v = 1$), $t=13.4$ ms (b3: $w = 6$, $N_v = 2$) and $t=28.0$ ms (b4: $w = 4$, $N_v = 4$).
(c1)-(c4) The 2D density profiles in the  $x-y$ plane of the superfluid with the obstacle, close to the instant the first vortex is emitted for $w_0=7$, corresponding to (c1) $t=9.6$ ms, (c2) $t=12.7$ ms, (c3) $t=13.4$ ms and (c4) $t=14.0$ ms.
(d) The time evolution of the local superfluid velocity (black line) and the local speed of sound (red line) extracted at $r_*$ for the imprinted circulation of  $w_0=6=w_c$ (i) and $w_0=7$ (ii). The vertical dashed grey lines for $w_0=7$   indicate the moments the vortices has crossed the low density channel, i.e. (c3) for the first emission.
} 
\label{Fig:theor_spiral_defect}
\end{figure*}
As the defect size is at the limit of the experimental resolution,  it is hard to know precisely its height. For this reason, we numerically study the decay of the persistent current for different defect heights close to the experimental $V_0 \approx \mu $ in the BEC superfluid, which correspond to $V_0/E_F \simeq 0.1$, and we find the value of $V_0/\mu=0.96$ to better fit both the experimental   $\langle w \rangle$ decay and the average number of emitted vortices. 
Simulations at different $V_0/\mu \in [0.9, 4]$ show that this parameter affects mainly $\Gamma $ and $N_v$ of the current decay, whereas $w_c$ barely depends on the obstacle height, changing only from $6$ to $5$ in the explored range. In the following we show only our studies for $V_0/\mu = 0.96$, relevant for the comparison with the experiment.   

Figure \ref{Fig:theor_spiral_defect}(a) shows the time evolution of the winding number $w$ and of the  number of emitted vortices $N_v$ for $w_0>w_c=6$, where $w_0=7, 8$ profiles are the ones shown in Fig.~\ref{Fig4}(a). All reported numerical data are fitted to provides the data reported as shaded region in Fig.~\ref{Fig4}(c-e).
In GPE simulations we find two critical circulation values: the first one, $w_c=6$, indicates the onset of dissipation, which happens through phase slippage caused by vortices entering from the inner edge of the ring. If we further increase $w_0$, we exceed a second critical value  $w_{c2}=8$ such that for $w_0>w_{c2}$ also antivortices (of opposite charge with respect to the winding number) could be emitted into the superfluid.
The interactions of vortices with antivortices, but also with other vortices or sound waves, cause annihilations or vortices/antivortices leaving the bulk, leading to the $N_v$ decrease in time. For $w_c<w_0<w_{c2}$ the vortex number remains instead constant to its maximum value.
Following the interferograms in time, as shown for $w_0=8$ in Fig.~\ref{Fig:theor_spiral_defect}(b1-b4), we note that as the total number of vortices increases, the winding number 
decreases in time by keeping their sum equal to $w_0$. The relation $N_v^{\mathrm tot} = w_0 -  w_F $ remains a good approximation even for higher $w_0$, as shown in Fig.~\ref{Fig4}(c). 
We note that the curves at $w_0>7$ are not monotonically decreasing in time, but they all present a unit increment of $w$ at $t \simeq 50 \,$ms, and those at $w_0>8$ also a second one at $t \simeq 100 \,$ms. Here, the numerical images show a vortex leaving the superfluid density towards the center of the ring, after the interaction with other vortices and density excitations, and a consequent increase of $w$ by one quantum. Such an effect happens only for high $w_0$, when a large number of vortices is emitted, and it is observable also in the darkest green dataset of Fig.~\ref{Fig4}(a). The experimental shot-to-shot fluctuations 
are 
too large to provide a conclusive demonstration of such vortex expulsion.

In order to give further insight into the microscopic description of the vortex emission process, we extract the 2D density of the superfluid in the $x-y$ plane and follow it in time around the instant when the first vortex is emitted, as shown in Fig.~\ref{Fig:theor_spiral_defect} for $w_{0}=7$. 
We find that for the considered defect height, vortices enter the superfluid from the inner edge of the ring, where its velocity is larger, and close to the defect position. In fact, a channel at low density is generated between the defect and the inner ring, which favours the entrance of the vortex from the very low density central region [Fig.~\ref{Fig:theor_spiral_defect} (c1, c2)]. Then this vortex propagates in the ring superfluid and after sometime another one is emitted. These vortices 
travel in a loop in the superfluid and after less than $100$ ms they return in the defect position, 
spiral around it and then continue their propagation, as shown in the movie provided as Supplemental Material \cite{SM}.
For $w_0>w_{c2}$, the vortex emission process becomes more complicated. 

We then extract the superfluid velocity field as $\mathbf{v}=\mathbf{j}/n$ where $\mathbf{j}$ is the current density and $n$ the superfluid density. We calculate its modulus value in the $x-y$ plane in the mid point between the defect and inner ring radius $r_*$ and we compare it to the local speed of sound $c=\sqrt{gn(r_*)/M}$.  The temporal profiles of both $v$ and $c$ are shown in Fig.~\ref{Fig:theor_spiral_defect}(d) for two examples of $w_0=6$ [(i)] and $w_0=7$ [(ii)]. The value of $c$ at $r_*$ is close to the bulk value $c_s$ of $5.8$ mm/s even though it varies in time due to the density exitations. We observe that in both cases there is an initial acceleration of the superfluid due to the constriction created between the defect and the inner edge of the ring. While for $w_0=6$ the acceleration stops almost immediately and $v$ always remains smaller than $c$, for $w_0=7$ instead the superfluid continues accelerating until it achieves a maximum value exceeding the local speed of sound. This correspond to the moment the linking low-density channel is created [Fig.~\ref{Fig:theor_spiral_defect}(c1)]. After that, a vortex enters the condensate, it causes a phase slip, and thus a jump in the superfluid velocity. The local minima of $v$ corresponds to the instants at which vortices depart from the region nearby the defect [Fig.~\ref{Fig:theor_spiral_defect}(c3)]. Thus, the superfluid velocity shows discrete jumps in its profile induced by phase slippage due to vortex emission, similar to the findings in superfluid helium \cite{hoskinson-2006,sato-packard-2012} and atomic Josephson junctions \cite{Xhani2020}.  
If we consider the initial superfluid velocity for $w_0=w_c=6$ at the ring inner edge as the critical value for vortex nucleation, we find it to be smaller than both the bulk speed of sound $c_s=5.8$ mm/s and the effective speed of sound obtained by averaging along the $z$-axis $c=4.74$ mm/s. 


Only for the first two peaks of Fig.~\ref{Fig:theor_spiral_defect}(d-ii) a vortex enters the bulk superfluid, while for the third one it does not have enough energy to do so. In fact, the former are characterized by a sharp jump following the vortex emission into the superfluid. 
The vortex emission shows a periodic character, and even for the more complicated emission of $w_0=10$,  $\Gamma$ scales linearly with $w_0-w_c$ as described in Fig.~\ref{Fig4}.
We note a small shift of a few ms in the vortex emission time between panels (a) and (d) of Fig.~\ref{Fig:theor_spiral_defect}, due to some differences in the boundaries conditions with [(a)] and without [(b)] the central disk \cite{SM}.

\section*{APPENDIX C: DENSITY PERTURBATIONS INTRODUCED BY THE PHASE IMPRINTING}

\begin{figure}[!t]
\centering
\vspace{11pt}
\includegraphics[width=1\columnwidth]{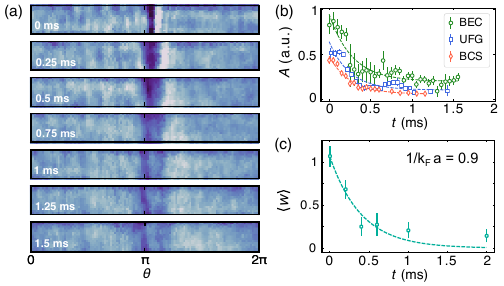}
\caption{Short-timescale dynamics after the imprinting. 
(a) Time evolution of the \textit{in situ} density profile of a BEC superfluid after a $\Delta \phi_I \simeq 2\pi$ imprinting. Each image consists of an average of $10$ absorption images, changed into polar coordinates. 
(b) Amplitude of the density depletion as a function of time after the suited imprinting pulse to populate the $w=1$ state in the three interaction regimes (see legend). Each data point corresponds to the fitted value and $1\sigma $ uncertainty from a gaussian fit of the integrated density. For UFG and BCS they are obtained from the analysis of the in situ superfluid density with no magnetic field sweep. Each curve is fitted with an exponential function providing a decay time of $\tau = \{0.28(5), 0.25(5), 0.25(3)\}$ ms in the BEC, UFG and BCS regime, respectively.
(c) Short-time dynamics of the average winding number and its SEM after an imprinting of $\Delta\phi_I = 2\pi$ in a strongly interacting superfluid at $1/k_F a \simeq 0.9$, measured from the interferograms acquired at the same interaction strength.
The dashed line represents an exponential fit of the data with $\tau = 400(80)$ $\mu$s.
}
\label{Fig:Density Depletion}
\end{figure}

As mentioned in the main text, the phase imprinting introduces density excitations in correspondence of the gradient discontinuity. To characterize them, we monitor the \textit{in situ} density profile of the ring superfluid at short times after the imprinting procedure. In particular, in correspondence of the maximum of the imprinting gradient, the atomic density is observed to present both a depletion and an accumulation of atoms close to each other [Fig.~\ref{Fig:Density Depletion}(a)], as a result of the action of the sharp antigradient in opposite direction. In fact, during the imprinting such an antigradient imparts a large momentum kick to the portion of atoms that it illuminates, creating both a density accumulation and depletion next to each other.
For increasing time after the imprinting, the density depletion and the accumulation are observed to travel in opposite direction with different velocities while progressively reducing their amplitude until they are not visible anymore [Fig.~\ref{Fig:Density Depletion}(a)]. 
We quantitatively characterize such behavior by fitting the integrated profile of the accumulation and depletion with two independent Gaussian functions. By analyzing the \textit{in situ} density profiles after the imprinting pulse suited to excite the $w = 1$ current state, we find that the accumulation travels at a velocity consistent with the speed of sound in each of the three interaction regimes, whereas the depletion proceedes at a fraction of $c_s$, i.e., at $\simeq \{ 50\%, 15\%, 40\%  \} \, c_s$ in BEC, UFG, BCS regimes, respectively. These values are consistent with a sound-like nature of the accumulation and a grey soliton-like nature of the depletion, despite the latter not necessarily possessing the soliton phase profile. In fact, the depletion is observed in the \textit{in situ} density profile after the imprinting of any $\Delta \phi_I \gtrsim \pi$ for all interaction regimes, but only for $\Delta \phi_I = \pi + 2n \pi$, with integer $n$, it is equipped by the phase profile proper of a soliton.
Both the sound and the soliton-like excitations are observed to exponentially reduce their amplitude to completely disappear from the density profile in a few ms timescale, as reported in Fig.~\ref{Fig:Density Depletion}(b) for the latter. 
A similar behavior is observed also in the GPE numerical simulation of the experimental imprinting profile \cite{SM}. In particular, when $\Delta \phi_I = \pi$, the decay dynamics of the density depletion resemble that observed for solitons \cite{Donadello2014, ku2016cascade}: it undergoes a snaking instability and eventually decays into a vortex/antivortex. For different imprinted phases, the depletion decay dynamics changes, but it always eventually produces a number of vortices that depend on the applied $\Delta \phi_I$. Those imprinting vortices are then observed to survive on top of the excited current without significantly perturbing it.

In order to understand the possible detrimental effect of these perturbations onto the superfluid circulation state, especially in the strongly interacting regime, which presents a shift on the $\langle w \rangle - \Delta \phi_I$ curve [Fig.~\ref{Fig3}(a)], we compare the density excitations dynamics with the the short-timescale evolution of the interferograms.
In the BEC superfluid, we observe that the spiral pattern of the fringes is present in the interferograms even at the shortest time after the imprinting we can probe, corresponding to the sole $1.2$ ms of the TOF, and that the $\langle w \rangle - \Delta \phi_I$ curve of Fig.~\ref{Fig3}(a) is independent on the time at which we probe. These observations demonstrate the negligible effect of the density excitations introduced by the imprinting on the current in the BEC regime.
On the other hand, the $50$ ms magnetic field ramp needed to acquire interferograms in the UFG and BCS regimes prevents us to explore the short-timescale dynamics of their interferograms. To investigate the role of the density excitations in strongly interacting superfluids, we therefore focus on the most strongly interacting BEC ($1/k_F a = 0.9$), that still shows clear interferograms with no need of sweeping the magnetic field and yet presents a sensible shift in the $\langle w \rangle - \Delta \phi_I$ curve [see Fig.~\ref{Fig3}(a) inset].
The time evolution of $\langle w \rangle$ after an imprinting of $\Delta \phi_I = 2\pi$ in such regime is reported in Fig.~\ref{Fig:Density Depletion}(c): the $2\pi$ phase winding is effectively imprinted, but it quickly decays in a few ms timescale, without giving rise to a persistent current. The comparable timescales of the phase winding decay and of the density excitations introduced by the imprinting hints to a connection between the two, as confirmed by the fact that when a barrier is added on the imprinting profile to avoid the density excitations, no decay is detected in the current after the imprinting of $2\pi$.
The polluting effect of the antigradient on the imprinted circulation produces a more dramatic effect on strongly interacting superfluids, that, as for the case of soliton imprinting \cite{Sacha2014}, show to be more sensitive to the sharpness of the imprinted profile. In future, it would be interesting to investigate further the effect of the imprinting profile on strongly interacting superfluids, both from an experimental and theoretical point of view.






\vspace*{-5pt}

%


\newpage

\setlength{\belowcaptionskip}{-5pt}
\renewcommand{\baselinestretch}{1.25}

\newcommand{\bra}[1]{\mbox{\ensuremath{\langle #1 \vert}}}
\newcommand{\ket}[1]{\mbox{\ensuremath{\vert #1 \rangle}}}
\newcommand{\Li}{$^{6}$Li }
\hyphenation{Fesh-bach}

\newcommand{\bcirc}{\textcolor{blue}{$\bigcirc$}}
\newcommand{\beq}{\begin{equation}}
\newcommand{\eeq}{\end{equation}}

\renewcommand{\thefigure}{S\arabic{figure}}
\setcounter{figure}{0}
\renewcommand{\theequation}{S.\arabic{equation}}
\setcounter{equation}{0}
\renewcommand{\thesection}{S.\arabic{section}}
\setcounter{section}{0}
\renewcommand{\thetable}{S\arabic{table}}
\setcounter{table}{0}

\renewcommand{\theHequation}{Supplement.\theequation}
\renewcommand{\theHfigure}{Supplement.\thefigure}
\renewcommand{\theHsection}{Supplement.\thesection}

\setlength{\tabcolsep}{18pt}

\onecolumngrid


\setcounter{equation}{0}
\setcounter{figure}{0}
\setcounter{table}{0}


\newpage
\begin{center}
\textbf{\large Supplemental Material\\[4mm] 
\Large Imprinting persistent currents in tunable fermionic rings}\\[4mm]
G. Del Pace,$^{1,\ast}$
K. Xhani,$^{1,}$
A. Muzi Falconi,$^{2, 1}$
M. Fedrizzi,$^{1}$ 
N. Grani,$^{1, 3}$
D. Hernandez Rajkov,$^{1}$
M. Inguscio,$^{4,1}$
F. Scazza$^{2, 1}$
W. J. Kwon$^{1}$
and
G. Roati$^{1}$
\\[2mm]
\emph{\small $^1$ Istituto Nazionale di Ottica del Consiglio Nazionale delle Ricerche (CNR-INO) and European Laboratory for Nonlinear Spectroscopy (LENS), University of Florence, 50019 Sesto Fiorentino, Italy}\\
\emph{\small $^2$ Department of Physics, University of Trieste, 34127 Trieste, Italy}\\
\emph{\small $^3$ Department of Physics and Astronomy, University of Florence, 50019 Sesto Fiorentino, Italy}\\
\emph{\small $^4$ Department of Engineering, Campus Bio-Medico University of Rome, 00128 Rome, Italy}
\end{center}
\vspace*{-10pt}
\begin{center}
\emph{\small ${}^\ast$ E-mail: delpace@lens.unifi.it}
\end{center}

\normalsize 

\setcounter{page}{1}

\section{Extracting local information from interference patterns}\label{Sec:InterfInfo}

\begin{figure}[h!]
\centering
\includegraphics[width=0.8\columnwidth]{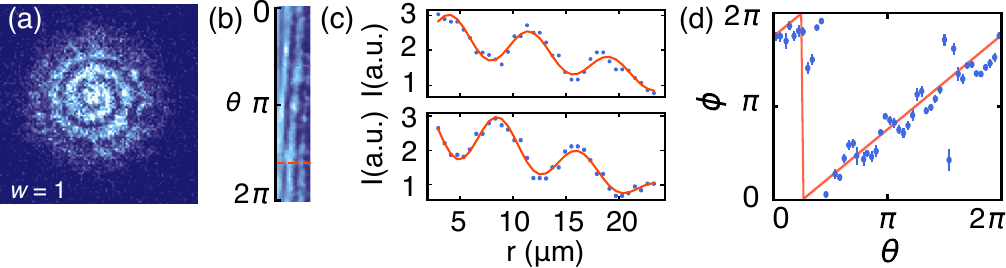}
\caption{Fitting procedure of the local phase $\phi(\theta)$. The acquired interferogram (a) is converted into polar coordinates, and a region of $1.5 \, \mu$m $<r < 12 \, \mu$m is selected (b). The $I(r, \theta)$ profile is sliced into $100$ strips, centred around a given azimuthal angle, and then fitted with a cosine function multiplied by a Gaussian envelope to extract the local phase shift $\phi(\theta)$ (c). The resulting $\phi (\theta)$ profile is fitted with a line to extract the winding number $w$ as slope (d), obtaining $w = 1.00(9)$. Symbols (errorbars) in panel (d) represents the weighted average (weighted standard deviation) over the fitted values of $\phi$ and their $1\sigma $ uncertainty in bins of $0.125\,$rad size.
}
\label{Fig: SM_Phase fitting}
\end{figure}

In this section we provide further information about the local fitting of the interferograms presented in the main text.
As the interference fringes have a high contrast in the BEC regime, to probe the winding number of UFG and BCS superfluids we acquire the interferograms after adiabatically sweeping the $s$-wave scattering length to the BEC regime with a $50$\,ms ramp before performing a time-of-flight (TOF) expansion.
%
%
Thanks to the high resolution $\sim 1\,\mu$m of our imaging system, we are able to measure the relative phase $\phi$ locally, with an azimuthal angular resolution of a few degrees. For this, we apply the following fitting procedure to the interferograms. We start by transforming the images from cartesian to polar coordinates, obtaining the interference pattern optical density $I$ as a function of the azimuthal angle $\theta$ and the radial position $r$ [see Fig.~\ref{Fig: SM_Phase fitting}(b)]. We then cut the $I(\theta,r)$ profile into slices at fixed $\theta$, and fit each $I(r)$ profile with a cosine multiplied by a Gaussian function [see Fig.~\ref{Fig: SM_Phase fitting}(c)]. The phase of each fitted cosine function allows to extract the local phase difference $\phi(\theta)$ between the reference disk and the ring. By performing a linear fit of $\phi(\theta)$ we extract the winding number $w$ [see Fig.~\ref{Fig: SM_Phase fitting}(d) and fitted $w$ values in the caption].
We verified that the described fitting procedure is applicable for any value of $w$ obtained in the experiments, although it becomes less reliable for $w \gtrsim 5$ due to the more complex pattern characterizing the interferograms.

\bigskip

\section{Obstacle characterization}

\begin{figure}[t]
\centering
\vspace{0pt}
\includegraphics[width=0.6\columnwidth]{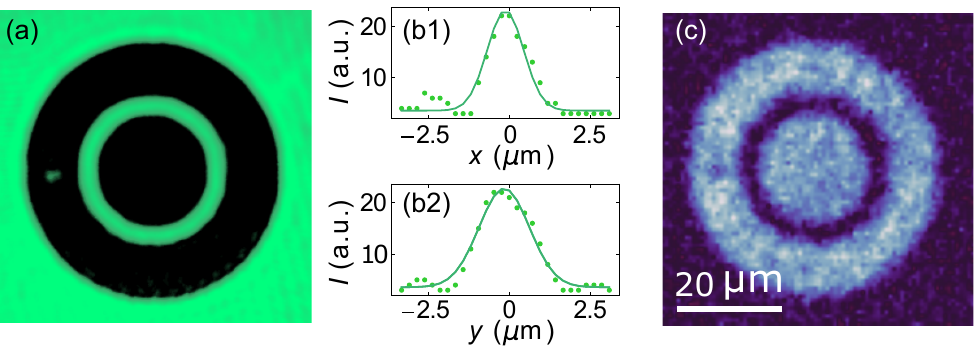}
\caption{Local defect characterization. (a) Image of the DMD pattern employed to introduce the small obstacle, visible as a small defect on the left side of the ring. (b1-2) Gaussian fit of the obstacle intensity profile $I$ along the $x-y$ direction. (c) \textit{In situ} image of a molecular BEC ring in presence of the local defect. The image is obtained as a single absorption image of the experimental sample.} 
\label{Fig SM: Obstacle}
\end{figure}

We characterize the small obstacle employed in Sec.~V of the main text by projecting the DMD-created image on a CCD camera
[Fig.~\ref{Fig SM: Obstacle}(a)] \cite{Kwon2020}. We perform a two-dimensional Gaussian fit of the obstacle intensity profile, finding that is has an approximately round profile of full width at half maximum (FWHM) of $1.6 \, \mu$m
[Figs.~\ref{Fig SM: Obstacle}(b1),(b2)], which is comparable with the characteristic correlation length of the investigated superfluids.  
From the amplitude of the fitted Gaussian profile we extract the potential height $V_0$, which is calibrated by employing the 
procedure described in Ref.~\citenum{Kwon2020}. The small size of the obstacle and the limited signal-to-noise ratio of CCD images make the calibration of $V_0$ not particularly precise. 
The obstacle height is adjusted to the value of $V_0/E_F \approx 0.1$ for the BEC superfluid, and to $V_0/E_F \approx 0.2$ for UFG and BCS ones by tuning the green light power impinging on the DMD. The obstacle height rescaled to the chemical potential in each regime is $V_0/\mu \simeq \{ 0.85, \, 0.19, \, 0.23\}$ for BEC, unitary and BCS superfluid, respectively.
Figure ~\ref{Fig SM: Obstacle}(c) shows the \textit{in situ} image of molecular BEC superfluid in presence of the local defect: the hole in the density produced by the latter confirms that $V_0/\mu \approx 1$.   

The same potential height calibration is used to measure the light shift $U_0$ induced by the light gradient employed in the imprinting procedure. With this knowledge, we compute the imprinted phase as $\Delta\phi_I = U_0\times t_I/\hbar$, where $t_I$ is the imprinting time. The error associated to this quantity in all figures corresponds to the $5$\% uncertainty on the calibration of $U_0$. 

\section{Maximum winding number state}\label{SM_Sec: w_max}

\begin{figure}[!t]
\centering
\vspace{11pt}
\includegraphics[width=0.65\columnwidth]{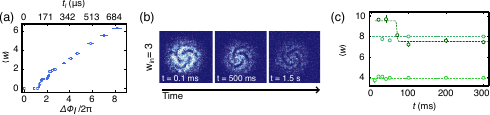}
\caption{(a) Average winding number as a function of the imprinted phase jump (lower x-axis) and the imprinting time (upper x-axis) in the unitary regime. Each point corresponds to the average of $\sim15$ experimental repetitions. Vertical error bars denote the standard deviation of the mean, horizontal ones are related only to the lower $x$-axis and are obtained considering the experimental uncertainty of $\sim 5$\% in the $U_0$ calibration.
(b) Time evolution of higher winding number in the BEC regime. The maximum stable circulation state that we can observe is $w=8$, while higher $w$ states decay back to this value. 
Each point corresponds to $\sim15$ experimental data points, error bars are the standard deviation of the mean. The dark green dashed line represents a sigmoidal fit of the data at $\langle w_0 \rangle \simeq 10$, lighter green lines correspond to $\langle w \rangle = 8, 4$.}
\label{Fig: SUP_Fig_wmax}
\end{figure}

By increasing the time duration of the imprinting pulse we are able to excite states with increasing winding number $w$, as reported in Fig.~\ref{Fig: SUP_Fig_wmax}(a). While for the shorter imprinting times we clearly observe a step-like profile of the populated circulation states, this profile becomes linear as we increase the imprinted phase jump.
This washing-out of the steps profile reflects the lower reliability of the imprinting procedure for the highest circulation states. We investigate the fidelity of our imprinting procedure for the $w=1$ state by acquiring more than 100 interferograms and evaluating the winding number for each of them. We estimate that the probability of populating the $w=1$ circulation state is $>99\%$ in the BEC and UFG regimes and a few percent lower in the BCS. This value decreases significantly if the imprinting time is increased
, with $w$ displaying significant shot-to-shot fluctuations.
In particular, we find that the maximum circulation state attainable with a single imprinting pulse of large $\Delta \phi _I$ is $w=6$ in all three superfluid regimes.
We ascribe this to the density excitations caused by the imprinting procedure, which become more dramatic for larger $t_I$ and could favour the entrance of vortices from the inner ring radius, effectively lowering the winding number in the ring.
As described in the main text, we thus typically employ multiple short imprinting pulses of $t_I \lesssim 200\,\mu$s to achieve the highest fidelity for exciting large-$w$ states.

Disregarding the short timescale phenomena described in the previous section, we find that for all $1 \geq w \geq w_{\tiny{\mbox{max}}}$ states the current is stable over timescales longer than that required to complete one entire loop around the ring. 
In particular, we find  $w_{\tiny{\mbox{max}}} = \{8, \, 9, \, 6 \}$ in the BEC, UFG and BCS regimes, respectively. Higher circulations $w >w_{\tiny{\mbox{max}}}$ are also accessible, but they decay to lower values over timescales comparable with a few round trip times, which is roughly $40\,$ms for $w=8$ [see Fig.~\ref{Fig: SUP_Fig_wmax}(b)]. At unitarity, to avoid the current decay during the magnetic field ramp of $50 \,$ms necessary to observe the interferograms in the strongly-interacting regime, we probe the states at $w>8$, i.e. $w_{\tiny{\mbox{max}}}$ in the BEC regime, with a different detection protocol. We employ a fast magnetic field ramp of $3.8\,$ms, timed such that during its last $1.3 \,$ms the atomic cloud perform a free TOF expansion. With such a detection protocol, the current at $w=9 $ is observed to persist for at leat $400 \,$ms, whereas the one at $w=10$ shows a similar decay to that plotted in Fig.~\ref{Fig: SUP_Fig_wmax}(b). We have checked that for currents at $w \leq 8$, the sweep duration does not affect the interferograms, as in this regime the critical current of the BEC superfluid is never exceeded. For all the investigated currents in the BCS regime, we employ the $50 \,$ms ramp, since $w_{\tiny{\mbox{max}}}$ in this regime is lower than in the BEC one, and longer sweeps provide higher contrast of the interferograms.
Consistently with our interpretation of $w_{\tiny{\mbox{max}}}$ as related to the current velocity at the inner ring radius exceeding the critical velocity for vortex nucleation, we have verified that the maximum winding strikingly depends on $R_{in}$, whereas it is completely unaffected by a change of $R_{out}$ as long as the ring thickness is kept larger than $\simeq 5 \, \mu$m. For thinner rings we observe progressively lower $w_{\tiny{\mbox{max}}}$ for a given $R_{in}$, most likely due to a more dramatic effect of the density excitations introduced by the phase imprinting, which could be more stable in more tightly confined geometries.


\section*{Theoretical description}

\subsection{GPE simulations of the experimental imprinting procedure}
The theoretical study in the BEC limit  is performed at $T=0$  by numerically solving the  time-dependent mean-field 3D Gross-Pitaevskii Equation (GPE):

\begin{equation}
i \hbar \frac{\partial \Psi (\bold{r},t)}{\partial t} = - \frac{\hbar^2}{2 M} \nabla^2 \Psi  (\bold{r},t) +  V \Psi (\bold{r},t) + g | \Psi (\bold{r},t) |^2 \Psi (\bold{r},t) \label{eq.GP.methods}
\end{equation}
where $\Psi(\bold{r},t)$ is the condensate wave function,  $M=2\, m$ the molecular mass, 
$V$ the external trapping potential, $g=4 \pi \hbar^2 a_M/M$ is the interaction strength with $a_M= 0.6 a \, = 53.3$ nm the molecular scattering length. 
In our study, we employ a ring-trap geometry with hard-wall boundaries such that the external potential is defined as follows: 
\begin{equation}
\displaystyle
V = \frac{1}{2} M \left ( \omega_\perp^2 r^2 + \omega_z^2 z^2 \right ) + V_{\rm ring} \; \; ,
\label{eq.GP.methods.potential}
\end{equation}
where $\{\omega_\perp , \, \omega_z \}= 2\pi \, \times \{2.5 \, , \, 396\}$ Hz
are the radial and axial trapping frequencies, respectively, and
\begin{equation}
\displaystyle
V_{\rm ring}=V_1 \left[\tanh\left(\frac{r-R_{\mathrm out}}{d}\right)+1\right]+V_1 \left[\tanh\left(\frac{R_{\rm in}-r}{d}\right)+1\right].
\label{Vbound}
\end{equation}
The parameters of $V_{\rm ring}$ are chosen such that the numerical density matches the experimental one, i.e.~$\{R_{\rm in} , \, R_{\rm out} \}=  \{9.6 \, , \, 21.0\} \, \mu \text{m}$ and $d=0.37 \, \mu \text{m}$, while $V_1 = 2.5 \, \mu$ is chosen to be larger than the chemical potential $\mu$ so the density goes to zero at the boundary. For a particle number equal to the experimental one $N=7.5\times 10^3$, we numerically obtain $\mu=1.06$ kHz and a healing length of $\xi = 0.61 \, \mu m$, consistent with the calculated ones for the experimental BEc superfluid.

Equation~\ref{eq.GP.methods} is solved numerically on a Cartesian grid of $\{ N_x , N_y , N_z \} = \{ 256 , 256 , 80\}$ points along $x$, $y$, and $z$ direction, respectively. We use a grid size of the same length in the radial plane, i.e.~$-34.846 \, \mu \text{m} \le x, y \le 34.846 \,  \mu \text{m}$ and $-11.0 \, \mu \text{m}  \le z \le 11.0 \, \mu \text{m}$. This gives rise to a grid spacing along the three directions of $\Delta x = \Delta y = 0.45 \, \xi$ and $\Delta z = 0.46 \, \xi$. The time step is instead set to $\Delta t = 1 \times 10^{-5}\, \omega_\perp^{-1}$.

\begin{figure*}[t!]
\centering
\vspace{11pt}
\includegraphics[width=.98\textwidth]{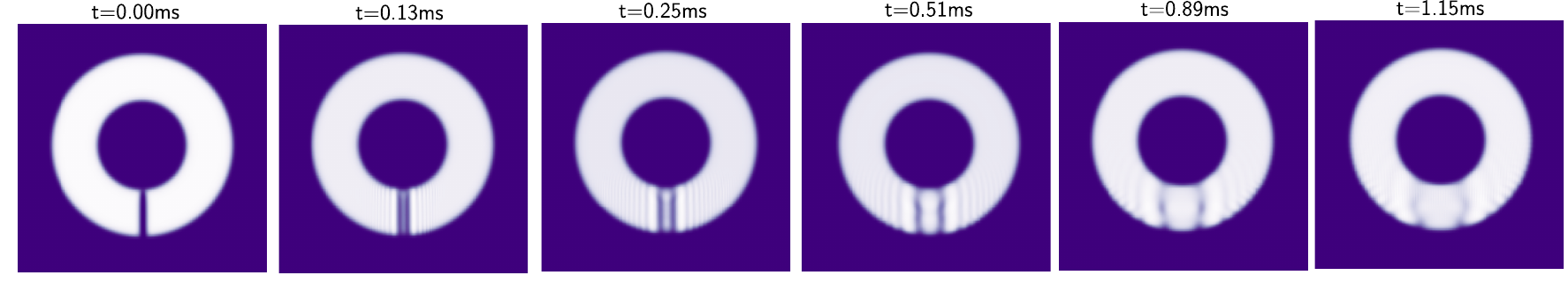}
\caption{Short-time evolution of the 2D density along the $x-y$ plane at $z=0$ showing the decay of the density depletion for an imprinted phase $\Delta \phi_I=2\pi$.}
\label{Fig:depl_decay}
\end{figure*}

The ground state wavefunction is found by solving the GPE in imaginary time. We then imprint a  phase  $\Delta \phi _I (\theta)=U(\theta) t_I/ \hbar$ on the initial condensate wavefunction, for a given imprinting time $t_I$. 
In  order to model the experimental imprinting potential shown in Fig.~2 of the main text, the imprinted function $U(\theta)$ is chosen to be  a decreasing function of the azimuthal angle $\theta$  in the range  $[0,2\pi  -\Delta \theta]$ starting from $U(\theta)=U_0$ at $\theta =0$ and reaching the zero value for $\theta=2\pi-\Delta \theta$. Then  it turns back to the initial value in $\Delta \theta = 0.15 $ rad. The presence of this opposite-sign gradient leads the superfluid density to present a cut after the imprinting, as shown in Fig.~\ref{Fig:depl_decay}(a1). 
This density depletion quickly decays over a few ms timescale into sound waves for $\Delta \phi_I / 2 \pi < 4$ [Fig.~\ref{Fig:depl_decay}], which does not affect the imprinted current. 
On the other hand, for $\Delta \phi_I / 2 \pi \geq 4$ 
the density excitations introduced by the imprinting give rise to vortices close to the inner radius of the ring, 
that initially perturb the current and then survive on top of it for long time. Furthermore, for $w_0\geq7$, the density excitation favours the vortex emission from the inner ring radius, setting the effective limit of $w$ reachable with a single imprinting pulse. 
This is consistent with the experimental findings of $w_{\mbox{\tiny{max}}}=6$ for a single imprinting pulse, as discussed in the main and in the previous section. 

In order to directly compare our numerical data with the experimental one, we employ the same interferometric technique for extracting the persistent current winding number.
As described in the main, it consists on adding a central disc condensate as phase reference and acquiring the interference pattern arising after a TOF expansion.
In order to do so, in our simulations we separate the ring from the disc by implementing a circular Gaussian barrier and changing $V_\mathrm{ring}$ of Eq.~\ref{Vbound} to:
\begin{equation}
\displaystyle
V_{\rm ring}=V_1 \left[\tanh\left(\frac{r-R_{\rm out}}{d} \right) +1 \right]+V_1 \exp \left[-\frac{2(R_{\rm in}-r)^2}{d^2} \right].
\label{Vbound2}
\end{equation}
where $d=1.1 \, \mu$m. The total number of pair considerend in the whole trap is now $10 \times 10^3$. 
To obtain the interferograms, we remove $V_\mathrm{ring}$ in $0.3$ ms, while keeping the harmonic vertical confinement. As the superfluid expands in the $x-y$ plane during the TOF, we modify accordingly the numerical grid to contain it all while keeping the same grid spacing as for the simple ring simulations. 
Thus, we  modify our grid size by taking the number of grid points to be $N_x=N_y=384$ for a grid size of length  $53.186 \mu \text{m} \le x, y \le 53.186 \mu \text{m}$, keeping so the grid spacing equal to the case where only ring superfluid is present.

\subsection{Numerical simulations of the persistent current decay}

The experimental studies of the decay of the persistent current in the presence of the defect are performed after the density depletion has already decayed. For this reason and in order to focus only in the effect of the defect on the supercurrent decay, in our numerical simulations of this part we simply imprint a phase in the initial condensate density, i.e. without including the antigradient potential.
In this way no density excitations perturb the superfluid initially. Moreover, the experiment is performed at finite temperature which means that the system is composed of condensate and thermal particles. The experimental condensed fraction decreases from $85\%$ before the imprinting to $80\%$ at the end of the time evolution in the presence of the defect. These values of condensed fraction are measured in the harmonic trap before lading the ring potential in the $x-y$ plane.
For the numerical study of the current decay we thus consider a number of particles equal to the $80\%$ of the total experimental $7.5 \times 10^3$ pairs, as with GPE only the condensate dynamic is taken into account.
By employing the collisionless Zaremba-Nikuni-Griffin model \cite{ZNG}, we have numerically checked that at the experimental temperature of $T=0.4 \, T_c$ and same condensate number \cite{Xhani2022}, the presence of a thermal cloud does not affect the critical winding $w_c$, but it rather alters both $\Gamma$ and $w_F$.

Also in this study, we initially investigate the ring superfluid only to describe the microscopic mechanism of vortex emission. Under these conditions, we extract the local superfluid velocity and the local speed of sound, as discussed in Appendix B of the main text. We then turn to the composite geometry including the disc condensate, in order to extract information about the winding number $w$ from the interferograms. 
We note that the experimental inner and outer ring radius in the presence of the inner disc are slightly different with respect to the simple ring, which give rise to smoother boundary conditions. For this reason we slightly change the parameters of $V_{\rm ring}$ of Eq.~\ref{Vbound2} by employing  $\{R_1 , \, R_2 \}=  \{10.2 \, , \, 20.6\} \, \mu \text{m}$ and $d=1.46\,  \mu  \text{m}$. With these new parameters the new chemical potential increases to $\mu =1.13$ kHz. The defect is parametrized with the following Gaussian shape:
\begin{equation}
    \displaystyle
V_{\rm defect}=V_0  \exp \left[-\frac{2(x-x_0)^2}{\sigma^2} \right] \exp \left[-\frac{2y^2}{\sigma^2} \right],
\end{equation}
with $x_0=-15.0 \, \mu$m, $\sigma=1.4\, \mu$m and $V_0=0.96 \, \mu$.
The studies performed in the superfluid ring with or without the inner disk condensate 
show a small shift on the timing of vortex nucleation, 
due to differences in the boundary conditions, 
but they display the same number of emitted vortices and the same critical circulation $w_c$.

\subsection{Thermodynamic properties of the superfluids throughout the BEC-BCS crossover}

In order to evaluate the thermodynamic properties of the superfluids in a unified manner throughout the BEC-BCS crossover
, we exploit the polytropic approximation for a trapped gas, where the effective polytropic index $\gamma = \partial \log \mu /\partial \log n$ 
leads to the power-law relation $\mu \propto n^{\gamma}$ \cite{Heiselberg2004,Haussmann2008}. In the BEC-limit, $\gamma=1$, while at unitarity and in the BCS-limit, $\gamma=2/3$. The polytropic approximation enables us to evaluate analytically $\mu$ and $E_F$ of the crossover superfluids when neglecting the shallow harmonic confinement in the radial direction 
and considering a hard-wall potential in the plane. These approximations are sustained by the fact that the radial Thomas-Fermi radius provided by the $2.5$ Hz harmonic trap in this direction is smaller than $R_\mathrm{out}$ and that the size of the box walls $d$ is comparable with the characteristic length of the superfluid in each regime. We verified that such approximation provides results in agreement within 1\% with the ones derived by accounting for the full trapping potential, i.e. by including also the radial harmonic confinement.

Under local density approximation, the Thomas-Fermi density profile is given by:
\begin{equation}
    n(r,z) = \left(\frac{\mu_0}{g_{\gamma}}\right)^{1/\gamma} \left[1 - \frac{z^2}{R_z^2} - \tilde{V}_{\mathrm{ring}}(r)\right]^{1/\gamma}
\label{Eq:density}
\end{equation}
where $R_z$ ($R_r$) are the Thomas-Fermi radius along the vertical (radial) direction, and $\tilde{V}_{\mathrm{ring}}(r)$ describe the hard wall potential:
\begin{equation}
    \tilde{V}_{\mathrm{ring}}(r) = 
    \begin{cases}
0 & \text{for $R_\mathrm{in} <r < R_\mathrm{out}$,} \\
\infty & \text{otherwise,}
\end{cases}
\end{equation}
where $R_\mathrm{in}$ and $R_\mathrm{out}$ are the internal and external ring radii.
We integrate the density of Eq.~\eqref{Eq:density} to obtain the total number of pairs $N$ and to calculate the corresponding chemical potential that reads:
\begin{equation}\label{eq:chemicalpotential}
\mu_0  =  \left[\frac{ \Gamma(\frac{1}{\gamma} +\frac{3}{2})}{\pi^{3/2}\Gamma(\frac{1}{\gamma}+1)} \frac{\sqrt{M/2}\omega_z N g_{\gamma}^{1/\gamma}}{R_\mathrm{out}^2-R_\mathrm{in}^2}\right]^{\frac{2\gamma}{\gamma+2}},
\end{equation}
where $\Gamma$ is the Gamma function and $g_{\gamma}$ is a prefactor that depends on the interaction limit: $g_\mathrm{BEC} = 4\pi\hbar^2a_M/M$ with $a_M = 0.6 \, a$ and $M = 2 \, m$ for $(k_Fa)^{-1}>1$, $g_\mathrm{BCS} = \frac{\hbar^2}{2m}(6\pi^2)^{2/3}$ for $(k_Fa_s)^{-1}<-1$ and $g_\mathrm{UFG} = \xi \frac{\hbar^2}{2m}(6\pi^2)^{2/3}$ for $-1<(k_Fa_s)^{-1}<1$, where $\xi$ is the Bertsch parameter taking the value of $0.37$ at unitarity at $T=0$ \cite{Haussmann2008}. 
Similarly, from the density of Eq.~\eqref{Eq:density} we calculate the Fermi energy:
\begin{equation}
E_F = 2\hbar\left[\frac{\hbar\omega_z N}{m(R_\mathrm{out}^2-R_\mathrm{in}^2)}\right]^{1/2}
\end{equation}

We apply the aforementioned relations to calculate $E_F$, $\mu$ and the speed of sound $c_s = \sqrt{\gamma \mu / m}$ for the three interaction regimes explored in this work, as reported in the main text. In particular, given the molecular nature of the BEC superfluid, we employ $M = 2 \, m$ for the sound speed calculation, yielding $\mu$ and $c_s$ values consistent with the results of the numerical simulation. On the other hand, given the small $|1/k_F a| $ value of the BCS superfluid, we calculate $\mu$ and $c_s$ in this regime by assuming the expression at unitarity to be valid in the strongly interacting regime of  $|1/k_F a| <1 $ and employing the Luttinger-Ward calculations across the BEC-BCS crossover of Ref.~\citenum{Haussmann2007} at $1/k_F a = -0.4$ \cite{Kwon2020}.
\vspace*{-10pt}


\begin{thebibliography}{70}%
\makeatletter
\providecommand \@ifxundefined [1]{%
 \@ifx{#1\undefined}
}%
\providecommand \@ifnum [1]{%
 \ifnum #1\expandafter \@firstoftwo
 \else \expandafter \@secondoftwo
 \fi
}%
\providecommand \@ifx [1]{%
 \ifx #1\expandafter \@firstoftwo
 \else \expandafter \@secondoftwo
 \fi
}%
\providecommand \natexlab [1]{#1}%
\providecommand \enquote  [1]{``#1''}%
\providecommand \bibnamefont  [1]{#1}%
\providecommand \bibfnamefont [1]{#1}%
\providecommand \citenamefont [1]{#1}%
\providecommand \href@noop [0]{\@secondoftwo}%
\providecommand \href [0]{\begingroup \@sanitize@url \@href}%
\providecommand \@href[1]{\@@startlink{#1}\@@href}%
\providecommand \@@href[1]{\endgroup#1\@@endlink}%
\providecommand \@sanitize@url [0]{\catcode `\\12\catcode `\$12\catcode
  `\&12\catcode `\#12\catcode `\^12\catcode `\_12\catcode `\%12\relax}%
\providecommand \@@startlink[1]{}%
\providecommand \@@endlink[0]{}%
\providecommand \url  [0]{\begingroup\@sanitize@url \@url }%
\providecommand \@url [1]{\endgroup\@href {#1}{\urlprefix }}%
\providecommand \urlprefix  [0]{URL }%
\providecommand \Eprint [0]{\href }%
\providecommand \doibase [0]{http://dx.doi.org/}%
\providecommand \selectlanguage [0]{\@gobble}%
\providecommand \bibinfo  [0]{\@secondoftwo}%
\providecommand \bibfield  [0]{\@secondoftwo}%
\providecommand \translation [1]{[#1]}%
\providecommand \BibitemOpen [0]{}%
\providecommand \bibitemStop [0]{}%
\providecommand \bibitemNoStop [0]{.\EOS\space}%
\providecommand \EOS [0]{\spacefactor3000\relax}%
\providecommand \BibitemShut  [1]{\csname bibitem#1\endcsname}%
\let\auto@bib@innerbib\@empty
\bibitem [{\citenamefont {Byers}\ and\ \citenamefont {Yang}(1961)}]{Byers1961}%
  \BibitemOpen
  \bibfield  {author} {\bibinfo {author} {\bibfnamefont {N.}~\bibnamefont
  {Byers}}\ and\ \bibinfo {author} {\bibfnamefont {C.~N.}\ \bibnamefont
  {Yang}},\ }\bibfield  {title} {\bibinfo {title} {\emph {Theoretical
  Considerations Concerning Quantized Magnetic Flux in Superconducting
  Cylinders}},\ }\href {\doibase 10.1103/PhysRevLett.7.46} {\bibfield
  {journal} {\bibinfo  {journal} {Phys. Rev. Lett.}\ }\textbf {\bibinfo
  {volume} {7}},\ \bibinfo {pages} {46} (\bibinfo {year} {1961})}\BibitemShut
  {NoStop}%
\bibitem [{\citenamefont {Bloch}(1965)}]{Bloch1965}%
  \BibitemOpen
  \bibfield  {author} {\bibinfo {author} {\bibfnamefont {F.}~\bibnamefont
  {Bloch}},\ }\bibfield  {title} {\bibinfo {title} {\emph {Off-Diagonal
  Long-Range Order and Persistent Currents in a Hollow Cylinder}},\ }\href
  {\doibase 10.1103/PhysRev.137.A787} {\bibfield  {journal} {\bibinfo
  {journal} {Phys. Rev.}\ }\textbf {\bibinfo {volume} {137}},\ \bibinfo {pages}
  {A787} (\bibinfo {year} {1965})}\BibitemShut {NoStop}%
\bibitem [{\citenamefont {Deaver}\ and\ \citenamefont
  {Fairbank}(1961)}]{Deaver1961}%
  \BibitemOpen
  \bibfield  {author} {\bibinfo {author} {\bibfnamefont {B.~S.}\ \bibnamefont
  {Deaver}}\ and\ \bibinfo {author} {\bibfnamefont {W.~M.}\ \bibnamefont
  {Fairbank}},\ }\bibfield  {title} {\bibinfo {title} {\emph {Experimental
  Evidence for Quantized Flux in Superconducting Cylinders}},\ }\href {\doibase
  10.1103/PhysRevLett.7.43} {\bibfield  {journal} {\bibinfo  {journal} {Phys.
  Rev. Lett.}\ }\textbf {\bibinfo {volume} {7}},\ \bibinfo {pages} {43}
  (\bibinfo {year} {1961})}\BibitemShut {NoStop}%
\bibitem [{\citenamefont {Doll}\ and\ \citenamefont
  {N\"abauer}(1961)}]{Doll1961}%
  \BibitemOpen
  \bibfield  {author} {\bibinfo {author} {\bibfnamefont {R.}~\bibnamefont
  {Doll}}\ and\ \bibinfo {author} {\bibfnamefont {M.}~\bibnamefont
  {N\"abauer}},\ }\bibfield  {title} {\bibinfo {title} {\emph {Experimental
  Proof of Magnetic Flux Quantization in a Superconducting Ring}},\ }\href
  {\doibase 10.1103/PhysRevLett.7.51} {\bibfield  {journal} {\bibinfo
  {journal} {Phys. Rev. Lett.}\ }\textbf {\bibinfo {volume} {7}},\ \bibinfo
  {pages} {51} (\bibinfo {year} {1961})}\BibitemShut {NoStop}%
\bibitem [{\citenamefont {Onsager}(1961)}]{Onsager1961}%
  \BibitemOpen
  \bibfield  {author} {\bibinfo {author} {\bibfnamefont {L.}~\bibnamefont
  {Onsager}},\ }\bibfield  {title} {\bibinfo {title} {\emph {Magnetic Flux
  Through a Superconducting Ring}},\ }\href {\doibase 10.1103/PhysRevLett.7.50}
  {\bibfield  {journal} {\bibinfo  {journal} {Phys. Rev. Lett.}\ }\textbf
  {\bibinfo {volume} {7}},\ \bibinfo {pages} {50} (\bibinfo {year}
  {1961})}\BibitemShut {NoStop}%
\bibitem [{\citenamefont {Bluhm}\ \emph {et~al.}(2009)\citenamefont {Bluhm},
  \citenamefont {Koshnick}, \citenamefont {Bert}, \citenamefont {Huber},\ and\
  \citenamefont {Moler}}]{Bluhm2009}%
  \BibitemOpen
  \bibfield  {author} {\bibinfo {author} {\bibfnamefont {H.}~\bibnamefont
  {Bluhm}}, \bibinfo {author} {\bibfnamefont {N.~C.}\ \bibnamefont {Koshnick}},
  \bibinfo {author} {\bibfnamefont {J.~A.}\ \bibnamefont {Bert}}, \bibinfo
  {author} {\bibfnamefont {M.~E.}\ \bibnamefont {Huber}}, \ and\ \bibinfo
  {author} {\bibfnamefont {K.~A.}\ \bibnamefont {Moler}},\ }\bibfield  {title}
  {\bibinfo {title} {\emph {Persistent Currents in Normal Metal Rings}},\
  }\href {\doibase 10.1103/PhysRevLett.102.136802} {\bibfield  {journal}
  {\bibinfo  {journal} {Phys. Rev. Lett.}\ }\textbf {\bibinfo {volume} {102}},\
  \bibinfo {pages} {136802} (\bibinfo {year} {2009})}\BibitemShut {NoStop}%
\bibitem [{\citenamefont {Bleszynski-Jayich}\ \emph {et~al.}(2009)\citenamefont
  {Bleszynski-Jayich}, \citenamefont {Shanks}, \citenamefont {Peaudecerf},
  \citenamefont {Ginossar}, \citenamefont {von Oppen}, \citenamefont
  {Glazman},\ and\ \citenamefont {Harris}}]{Bleszynski2009}%
  \BibitemOpen
  \bibfield  {author} {\bibinfo {author} {\bibfnamefont {A.~C.}\ \bibnamefont
  {Bleszynski-Jayich}}, \bibinfo {author} {\bibfnamefont {W.~E.}\ \bibnamefont
  {Shanks}}, \bibinfo {author} {\bibfnamefont {B.}~\bibnamefont {Peaudecerf}},
  \bibinfo {author} {\bibfnamefont {E.}~\bibnamefont {Ginossar}}, \bibinfo
  {author} {\bibfnamefont {F.}~\bibnamefont {von Oppen}}, \bibinfo {author}
  {\bibfnamefont {L.}~\bibnamefont {Glazman}}, \ and\ \bibinfo {author}
  {\bibfnamefont {J.~G.~E.}\ \bibnamefont {Harris}},\ }\bibfield  {title}
  {\bibinfo {title} {\emph {Persistent Currents in Normal Metal Rings}},\
  }\href {\doibase 10.1126/science.1178139} {\bibfield  {journal} {\bibinfo
  {journal} {Science}\ }\textbf {\bibinfo {volume} {326}},\ \bibinfo {pages}
  {272} (\bibinfo {year} {2009})}\BibitemShut {NoStop}%
\bibitem [{\citenamefont {Tinkham}(2004)}]{TinkhamBook}%
  \BibitemOpen
  \bibfield  {author} {\bibinfo {author} {\bibfnamefont {M.}~\bibnamefont
  {Tinkham}},\ }\href@noop {} {\emph {\bibinfo {title} {Introduction to
  superconductivity}}}\ (\bibinfo  {publisher} {Courier Corporation},\ \bibinfo
  {year} {2004})\BibitemShut {NoStop}%
\bibitem [{\citenamefont {Dalibard}\ \emph {et~al.}(2011)\citenamefont
  {Dalibard}, \citenamefont {Gerbier}, \citenamefont {Juzeliunas},\ and\
  \citenamefont {\"Ohberg}}]{Dalibard2011}%
  \BibitemOpen
  \bibfield  {author} {\bibinfo {author} {\bibfnamefont {J.}~\bibnamefont
  {Dalibard}}, \bibinfo {author} {\bibfnamefont {F.}~\bibnamefont {Gerbier}},
  \bibinfo {author} {\bibfnamefont {G.}~\bibnamefont {Juzeliunas}}, \ and\
  \bibinfo {author} {\bibfnamefont {P.}~\bibnamefont {\"Ohberg}},\ }\bibfield
  {title} {\bibinfo {title} {\emph {Colloquium: Artificial gauge potentials for
  neutral atoms}},\ }\href {\doibase 10.1103/RevModPhys.83.1523} {\bibfield
  {journal} {\bibinfo  {journal} {Rev. Mod. Phys.}\ }\textbf {\bibinfo {volume}
  {83}},\ \bibinfo {pages} {1523} (\bibinfo {year} {2011})}\BibitemShut
  {NoStop}%
\bibitem [{\citenamefont {Goldman}\ \emph {et~al.}(2014)\citenamefont
  {Goldman}, \citenamefont {Juzeli{\=u}nas}, \citenamefont {{\"O}hberg},\ and\
  \citenamefont {Spielman}}]{Goldman2014}%
  \BibitemOpen
  \bibfield  {author} {\bibinfo {author} {\bibfnamefont {N.}~\bibnamefont
  {Goldman}}, \bibinfo {author} {\bibfnamefont {G.}~\bibnamefont
  {Juzeli{\=u}nas}}, \bibinfo {author} {\bibfnamefont {P.}~\bibnamefont
  {{\"O}hberg}}, \ and\ \bibinfo {author} {\bibfnamefont {I.~B.}\ \bibnamefont
  {Spielman}},\ }\bibfield  {title} {\bibinfo {title} {\emph {Light-induced
  gauge fields for ultracold atoms}},\ }\href@noop {} {\bibfield  {journal}
  {\bibinfo  {journal} {Rep. Progr. Phys.}\ }\textbf {\bibinfo {volume} {77}},\
  \bibinfo {pages} {126401} (\bibinfo {year} {2014})}\BibitemShut {NoStop}%
\bibitem [{\citenamefont {Aharonov}\ and\ \citenamefont
  {Bohm}(1959)}]{Aharonov1959}%
  \BibitemOpen
  \bibfield  {author} {\bibinfo {author} {\bibfnamefont {Y.}~\bibnamefont
  {Aharonov}}\ and\ \bibinfo {author} {\bibfnamefont {D.}~\bibnamefont
  {Bohm}},\ }\bibfield  {title} {\bibinfo {title} {\emph {Significance of
  electromagnetic potentials in the quantum theory}},\ }\href@noop {}
  {\bibfield  {journal} {\bibinfo  {journal} {Phys. Rev.}\ }\textbf {\bibinfo
  {volume} {115}},\ \bibinfo {pages} {485} (\bibinfo {year}
  {1959})}\BibitemShut {NoStop}%
\bibitem [{\citenamefont {Kadin}(1999)}]{Kadin1999book}%
  \BibitemOpen
  \bibfield  {author} {\bibinfo {author} {\bibfnamefont {A.}~\bibnamefont
  {Kadin}},\ }\href@noop {} {\emph {\bibinfo {title} {Introduction to
  Superconducting Circuits}}}\ (\bibinfo  {publisher} {Wiley},\ \bibinfo {year}
  {1999})\BibitemShut {NoStop}%
\bibitem [{\citenamefont {Leggett}(1999)}]{Leggett1999}%
  \BibitemOpen
  \bibfield  {author} {\bibinfo {author} {\bibfnamefont {A.~J.}\ \bibnamefont
  {Leggett}},\ }\bibfield  {title} {\bibinfo {title} {\emph {Superfluidity}},\
  }\href {\doibase 10.1103/RevModPhys.71.S318} {\bibfield  {journal} {\bibinfo
  {journal} {Rev. Mod. Phys.}\ }\textbf {\bibinfo {volume} {71}},\ \bibinfo
  {pages} {S318} (\bibinfo {year} {1999})}\BibitemShut {NoStop}%
\bibitem [{\citenamefont {Mueller}(2002)}]{Mueller2002}%
  \BibitemOpen
  \bibfield  {author} {\bibinfo {author} {\bibfnamefont {E.~J.}\ \bibnamefont
  {Mueller}},\ }\bibfield  {title} {\bibinfo {title} {\emph {Superfluidity and
  mean-field energy loops: Hysteretic behavior in Bose-Einstein condensates}},\
  }\href {\doibase 10.1103/PhysRevA.66.063603} {\bibfield  {journal} {\bibinfo
  {journal} {Phys. Rev. A}\ }\textbf {\bibinfo {volume} {66}},\ \bibinfo
  {pages} {063603} (\bibinfo {year} {2002})}\BibitemShut {NoStop}%
\bibitem [{\citenamefont {Bloch}(1973)}]{Bloch1975}%
  \BibitemOpen
  \bibfield  {author} {\bibinfo {author} {\bibfnamefont {F.}~\bibnamefont
  {Bloch}},\ }\bibfield  {title} {\bibinfo {title} {\emph {Superfluidity in a
  Ring}},\ }\href {\doibase 10.1103/PhysRevA.7.2187} {\bibfield  {journal}
  {\bibinfo  {journal} {Phys. Rev. A}\ }\textbf {\bibinfo {volume} {7}},\
  \bibinfo {pages} {2187} (\bibinfo {year} {1973})}\BibitemShut {NoStop}%
\bibitem [{\citenamefont {Ryu}\ \emph {et~al.}(2007)\citenamefont {Ryu},
  \citenamefont {Andersen}, \citenamefont {Clade}, \citenamefont {Natarajan},
  \citenamefont {Helmerson},\ and\ \citenamefont {Phillips}}]{Ryu2007}%
  \BibitemOpen
  \bibfield  {author} {\bibinfo {author} {\bibfnamefont {C.}~\bibnamefont
  {Ryu}}, \bibinfo {author} {\bibfnamefont {M.~F.}\ \bibnamefont {Andersen}},
  \bibinfo {author} {\bibfnamefont {P.}~\bibnamefont {Clade}}, \bibinfo
  {author} {\bibfnamefont {V.}~\bibnamefont {Natarajan}}, \bibinfo {author}
  {\bibfnamefont {K.}~\bibnamefont {Helmerson}}, \ and\ \bibinfo {author}
  {\bibfnamefont {W.~D.}\ \bibnamefont {Phillips}},\ }\bibfield  {title}
  {\bibinfo {title} {\emph {Observation of persistent flow of a Bose-Einstein
  condensate in a toroidal trap}},\ }\href@noop {} {\bibfield  {journal}
  {\bibinfo  {journal} {Phys. Rev. Lett.}\ }\textbf {\bibinfo {volume} {99}},\
  \bibinfo {pages} {260401} (\bibinfo {year} {2007})}\BibitemShut {NoStop}%
\bibitem [{\citenamefont {Moulder}\ \emph {et~al.}(2012)\citenamefont
  {Moulder}, \citenamefont {Beattie}, \citenamefont {Smith}, \citenamefont
  {Tammuz},\ and\ \citenamefont {Hadzibabic}}]{Moulder2012}%
  \BibitemOpen
  \bibfield  {author} {\bibinfo {author} {\bibfnamefont {S.}~\bibnamefont
  {Moulder}}, \bibinfo {author} {\bibfnamefont {S.}~\bibnamefont {Beattie}},
  \bibinfo {author} {\bibfnamefont {R.~P.}\ \bibnamefont {Smith}}, \bibinfo
  {author} {\bibfnamefont {N.}~\bibnamefont {Tammuz}}, \ and\ \bibinfo {author}
  {\bibfnamefont {Z.}~\bibnamefont {Hadzibabic}},\ }\bibfield  {title}
  {\bibinfo {title} {\emph {Quantized supercurrent decay in an annular
  Bose-Einstein condensate}},\ }\href@noop {} {\bibfield  {journal} {\bibinfo
  {journal} {Phys. Rev. A}\ }\textbf {\bibinfo {volume} {86}},\ \bibinfo
  {pages} {013629} (\bibinfo {year} {2012})}\BibitemShut {NoStop}%
\bibitem [{\citenamefont {Beattie}\ \emph {et~al.}(2013)\citenamefont
  {Beattie}, \citenamefont {Moulder}, \citenamefont {Fletcher},\ and\
  \citenamefont {Hadzibabic}}]{Beattie2013}%
  \BibitemOpen
  \bibfield  {author} {\bibinfo {author} {\bibfnamefont {S.}~\bibnamefont
  {Beattie}}, \bibinfo {author} {\bibfnamefont {S.}~\bibnamefont {Moulder}},
  \bibinfo {author} {\bibfnamefont {R.~J.}\ \bibnamefont {Fletcher}}, \ and\
  \bibinfo {author} {\bibfnamefont {Z.}~\bibnamefont {Hadzibabic}},\ }\bibfield
   {title} {\bibinfo {title} {\emph {Persistent currents in spinor
  condensates}},\ }\href@noop {} {\bibfield  {journal} {\bibinfo  {journal}
  {Phys. Rev. Lett.}\ }\textbf {\bibinfo {volume} {110}},\ \bibinfo {pages}
  {025301} (\bibinfo {year} {2013})}\BibitemShut {NoStop}%
\bibitem [{\citenamefont {Wright}\ \emph {et~al.}(2013)\citenamefont {Wright},
  \citenamefont {Blakestad}, \citenamefont {Lobb}, \citenamefont {Phillips},\
  and\ \citenamefont {Campbell}}]{Wright2013_driving}%
  \BibitemOpen
  \bibfield  {author} {\bibinfo {author} {\bibfnamefont {K.~C.}\ \bibnamefont
  {Wright}}, \bibinfo {author} {\bibfnamefont {R.~B.}\ \bibnamefont
  {Blakestad}}, \bibinfo {author} {\bibfnamefont {C.~J.}\ \bibnamefont {Lobb}},
  \bibinfo {author} {\bibfnamefont {W.~D.}\ \bibnamefont {Phillips}}, \ and\
  \bibinfo {author} {\bibfnamefont {G.~K.}\ \bibnamefont {Campbell}},\
  }\bibfield  {title} {\bibinfo {title} {\emph {Driving phase slips in a
  superfluid atom circuit with a rotating weak link}},\ }\href@noop {}
  {\bibfield  {journal} {\bibinfo  {journal} {Phys. Rev. Lett.}\ }\textbf
  {\bibinfo {volume} {110}},\ \bibinfo {pages} {025302} (\bibinfo {year}
  {2013})}\BibitemShut {NoStop}%
\bibitem [{\citenamefont {Kumar}\ \emph {et~al.}(2017)\citenamefont {Kumar},
  \citenamefont {Eckel}, \citenamefont {Jendrzejewski},\ and\ \citenamefont
  {Campbell}}]{Kumar2017}%
  \BibitemOpen
  \bibfield  {author} {\bibinfo {author} {\bibfnamefont {A.}~\bibnamefont
  {Kumar}}, \bibinfo {author} {\bibfnamefont {S.}~\bibnamefont {Eckel}},
  \bibinfo {author} {\bibfnamefont {F.}~\bibnamefont {Jendrzejewski}}, \ and\
  \bibinfo {author} {\bibfnamefont {G.~K.}\ \bibnamefont {Campbell}},\
  }\bibfield  {title} {\bibinfo {title} {\emph {Temperature-induced decay of
  persistent currents in a superfluid ultracold gas}},\ }\href@noop {}
  {\bibfield  {journal} {\bibinfo  {journal} {Phys. Rev. A}\ }\textbf {\bibinfo
  {volume} {95}},\ \bibinfo {pages} {021602(R)} (\bibinfo {year}
  {2017})}\BibitemShut {NoStop}%
\bibitem [{\citenamefont {Cai}\ \emph {et~al.}(2022)\citenamefont {Cai},
  \citenamefont {Allman}, \citenamefont {Sabharwal},\ and\ \citenamefont
  {Wright}}]{Cai2021}%
  \BibitemOpen
  \bibfield  {author} {\bibinfo {author} {\bibfnamefont {Y.}~\bibnamefont
  {Cai}}, \bibinfo {author} {\bibfnamefont {D.~G.}\ \bibnamefont {Allman}},
  \bibinfo {author} {\bibfnamefont {P.}~\bibnamefont {Sabharwal}}, \ and\
  \bibinfo {author} {\bibfnamefont {K.~C.}\ \bibnamefont {Wright}},\ }\bibfield
   {title} {\bibinfo {title} {\emph {Persistent currents in rings of ultracold
  fermionic atoms}},\ }\href@noop {} {\bibfield  {journal} {\bibinfo  {journal}
  {Physical Review Letters}\ }\textbf {\bibinfo {volume} {128}},\ \bibinfo
  {pages} {150401} (\bibinfo {year} {2022})}\BibitemShut {NoStop}%
\bibitem [{\citenamefont {Dubessy}\ \emph {et~al.}(2012)\citenamefont
  {Dubessy}, \citenamefont {Liennard}, \citenamefont {Pedri},\ and\
  \citenamefont {Perrin}}]{Dubessy2012}%
  \BibitemOpen
  \bibfield  {author} {\bibinfo {author} {\bibfnamefont {R.}~\bibnamefont
  {Dubessy}}, \bibinfo {author} {\bibfnamefont {T.}~\bibnamefont {Liennard}},
  \bibinfo {author} {\bibfnamefont {P.}~\bibnamefont {Pedri}}, \ and\ \bibinfo
  {author} {\bibfnamefont {H.}~\bibnamefont {Perrin}},\ }\bibfield  {title}
  {\bibinfo {title} {\emph {Critical rotation of an annular superfluid
  Bose-Einstein condensate}},\ }\href {\doibase 10.1103/PhysRevA.86.011602}
  {\bibfield  {journal} {\bibinfo  {journal} {Phys. Rev. A}\ }\textbf {\bibinfo
  {volume} {86}},\ \bibinfo {pages} {011602(R)} (\bibinfo {year}
  {2012})}\BibitemShut {NoStop}%
\bibitem [{\citenamefont {Polo}\ \emph {et~al.}(2019)\citenamefont {Polo},
  \citenamefont {Dubessy}, \citenamefont {Pedri}, \citenamefont {Perrin},\ and\
  \citenamefont {Minguzzi}}]{Polo2019}%
  \BibitemOpen
  \bibfield  {author} {\bibinfo {author} {\bibfnamefont {J.}~\bibnamefont
  {Polo}}, \bibinfo {author} {\bibfnamefont {R.}~\bibnamefont {Dubessy}},
  \bibinfo {author} {\bibfnamefont {P.}~\bibnamefont {Pedri}}, \bibinfo
  {author} {\bibfnamefont {H.}~\bibnamefont {Perrin}}, \ and\ \bibinfo {author}
  {\bibfnamefont {A.}~\bibnamefont {Minguzzi}},\ }\bibfield  {title} {\bibinfo
  {title} {\emph {Oscillations and Decay of Superfluid Currents in a
  One-Dimensional Bose Gas on a Ring}},\ }\href {\doibase
  10.1103/PhysRevLett.123.195301} {\bibfield  {journal} {\bibinfo  {journal}
  {Phys. Rev. Lett.}\ }\textbf {\bibinfo {volume} {123}},\ \bibinfo {pages}
  {195301} (\bibinfo {year} {2019})}\BibitemShut {NoStop}%
\bibitem [{\citenamefont {Kunimi}\ and\ \citenamefont
  {Danshita}(2019)}]{Kunimi2019}%
  \BibitemOpen
  \bibfield  {author} {\bibinfo {author} {\bibfnamefont {M.}~\bibnamefont
  {Kunimi}}\ and\ \bibinfo {author} {\bibfnamefont {I.}~\bibnamefont
  {Danshita}},\ }\bibfield  {title} {\bibinfo {title} {\emph {Decay mechanisms
  of superflow of Bose-Einstein condensates in ring traps}},\ }\href {\doibase
  10.1103/PhysRevA.99.043613} {\bibfield  {journal} {\bibinfo  {journal} {Phys.
  Rev. A}\ }\textbf {\bibinfo {volume} {99}},\ \bibinfo {pages} {043613}
  (\bibinfo {year} {2019})}\BibitemShut {NoStop}%
\bibitem [{\citenamefont {Mehdi}\ \emph {et~al.}(2021)\citenamefont {Mehdi},
  \citenamefont {Bradley}, \citenamefont {Hope},\ and\ \citenamefont
  {Szigeti}}]{Mehdi2021}%
  \BibitemOpen
  \bibfield  {author} {\bibinfo {author} {\bibfnamefont {Z.}~\bibnamefont
  {Mehdi}}, \bibinfo {author} {\bibfnamefont {A.~S.}\ \bibnamefont {Bradley}},
  \bibinfo {author} {\bibfnamefont {J.~J.}\ \bibnamefont {Hope}}, \ and\
  \bibinfo {author} {\bibfnamefont {S.~S.}\ \bibnamefont {Szigeti}},\
  }\bibfield  {title} {\bibinfo {title} {\emph {{Superflow decay in a toroidal
  Bose gas: The effect of quantum and thermal fluctuations}}},\ }\href
  {\doibase 10.21468/SciPostPhys.11.4.080} {\bibfield  {journal} {\bibinfo
  {journal} {SciPost Phys.}\ }\textbf {\bibinfo {volume} {11}},\ \bibinfo
  {pages} {80} (\bibinfo {year} {2021})}\BibitemShut {NoStop}%
\bibitem [{\citenamefont {Ferrero}\ \emph {et~al.}(2021)\citenamefont
  {Ferrero}, \citenamefont {Foini}, \citenamefont {Giamarchi}, \citenamefont
  {Kolton},\ and\ \citenamefont {Rosso}}]{Ferrero2021}%
  \BibitemOpen
  \bibfield  {author} {\bibinfo {author} {\bibfnamefont {E.~E.}\ \bibnamefont
  {Ferrero}}, \bibinfo {author} {\bibfnamefont {L.}~\bibnamefont {Foini}},
  \bibinfo {author} {\bibfnamefont {T.}~\bibnamefont {Giamarchi}}, \bibinfo
  {author} {\bibfnamefont {A.~B.}\ \bibnamefont {Kolton}}, \ and\ \bibinfo
  {author} {\bibfnamefont {A.}~\bibnamefont {Rosso}},\ }\bibfield  {title}
  {\bibinfo {title} {\emph {Creep Motion of Elastic Interfaces Driven in a
  Disordered Landscape}},\ }\href {\doibase
  10.1146/annurev-conmatphys-031119-050725} {\bibfield  {journal} {\bibinfo
  {journal} {Annu. Rev. Cond. Mat. Phys.}\ }\textbf {\bibinfo {volume} {12}},\
  \bibinfo {pages} {111} (\bibinfo {year} {2021})}\BibitemShut {NoStop}%
\bibitem [{\citenamefont {Amico}\ \emph {et~al.}(2021)\citenamefont {Amico},
  \citenamefont {Boshier}, \citenamefont {Birkl}, \citenamefont {Minguzzi},
  \citenamefont {Miniatura}, \citenamefont {Kwek}, \citenamefont {Aghamalyan},
  \citenamefont {Ahufinger}, \citenamefont {Anderson}, \citenamefont {Andrei}
  \emph {et~al.}}]{Amico2021}%
  \BibitemOpen
  \bibfield  {author} {\bibinfo {author} {\bibfnamefont {L.}~\bibnamefont
  {Amico}}, \bibinfo {author} {\bibfnamefont {M.}~\bibnamefont {Boshier}},
  \bibinfo {author} {\bibfnamefont {G.}~\bibnamefont {Birkl}}, \bibinfo
  {author} {\bibfnamefont {A.}~\bibnamefont {Minguzzi}}, \bibinfo {author}
  {\bibfnamefont {C.}~\bibnamefont {Miniatura}}, \bibinfo {author}
  {\bibfnamefont {L.-C.}\ \bibnamefont {Kwek}}, \bibinfo {author}
  {\bibfnamefont {D.}~\bibnamefont {Aghamalyan}}, \bibinfo {author}
  {\bibfnamefont {V.}~\bibnamefont {Ahufinger}}, \bibinfo {author}
  {\bibfnamefont {D.}~\bibnamefont {Anderson}}, \bibinfo {author}
  {\bibfnamefont {N.}~\bibnamefont {Andrei}},  \emph {et~al.},\ }\bibfield
  {title} {\bibinfo {title} {\emph {Roadmap on Atomtronics: State of the art
  and perspective}},\ }\href@noop {} {\bibfield  {journal} {\bibinfo  {journal}
  {AVS Quantum Science}\ }\textbf {\bibinfo {volume} {3}},\ \bibinfo {pages}
  {039201} (\bibinfo {year} {2021})}\BibitemShut {NoStop}%
\bibitem [{\citenamefont {Amico}\ \emph {et~al.}(2022)\citenamefont {Amico},
  \citenamefont {Anderson}, \citenamefont {Boshier}, \citenamefont {Brantut},
  \citenamefont {Kwek}, \citenamefont {Minguzzi},\ and\ \citenamefont {von
  Klitzing}}]{Amico2021_atomtronic}%
  \BibitemOpen
  \bibfield  {author} {\bibinfo {author} {\bibfnamefont {L.}~\bibnamefont
  {Amico}}, \bibinfo {author} {\bibfnamefont {D.}~\bibnamefont {Anderson}},
  \bibinfo {author} {\bibfnamefont {M.}~\bibnamefont {Boshier}}, \bibinfo
  {author} {\bibfnamefont {J.-P.}\ \bibnamefont {Brantut}}, \bibinfo {author}
  {\bibfnamefont {L.-C.}\ \bibnamefont {Kwek}}, \bibinfo {author}
  {\bibfnamefont {A.}~\bibnamefont {Minguzzi}}, \ and\ \bibinfo {author}
  {\bibfnamefont {W.}~\bibnamefont {von Klitzing}},\ }\bibfield  {title}
  {\bibinfo {title} {\emph {Colloquium: Atomtronic circuits: From many-body
  physics to quantum technologies}},\ }\href {\doibase
  10.1103/RevModPhys.94.041001} {\bibfield  {journal} {\bibinfo  {journal}
  {Rev. Mod. Phys.}\ }\textbf {\bibinfo {volume} {94}},\ \bibinfo {pages}
  {041001} (\bibinfo {year} {2022})}\BibitemShut {NoStop}%
\bibitem [{\citenamefont {Aghamalyan}\ \emph {et~al.}(2015)\citenamefont
  {Aghamalyan}, \citenamefont {Cominotti}, \citenamefont {Rizzi}, \citenamefont
  {Rossini}, \citenamefont {Hekking}, \citenamefont {Minguzzi}, \citenamefont
  {Kwek},\ and\ \citenamefont {Amico}}]{Aghamalyan2015}%
  \BibitemOpen
  \bibfield  {author} {\bibinfo {author} {\bibfnamefont {D.}~\bibnamefont
  {Aghamalyan}}, \bibinfo {author} {\bibfnamefont {M.}~\bibnamefont
  {Cominotti}}, \bibinfo {author} {\bibfnamefont {M.}~\bibnamefont {Rizzi}},
  \bibinfo {author} {\bibfnamefont {D.}~\bibnamefont {Rossini}}, \bibinfo
  {author} {\bibfnamefont {F.}~\bibnamefont {Hekking}}, \bibinfo {author}
  {\bibfnamefont {A.}~\bibnamefont {Minguzzi}}, \bibinfo {author}
  {\bibfnamefont {L.-C.}\ \bibnamefont {Kwek}}, \ and\ \bibinfo {author}
  {\bibfnamefont {L.}~\bibnamefont {Amico}},\ }\bibfield  {title} {\bibinfo
  {title} {\emph {Coherent superposition of current flows in an atomtronic
  quantum interference device}},\ }\href@noop {} {\bibfield  {journal}
  {\bibinfo  {journal} {New J. Phys.}\ }\textbf {\bibinfo {volume} {17}},\
  \bibinfo {pages} {045023} (\bibinfo {year} {2015})}\BibitemShut {NoStop}%
\bibitem [{\citenamefont {Ramanathan}\ \emph {et~al.}(2011)\citenamefont
  {Ramanathan}, \citenamefont {Wright}, \citenamefont {Muniz}, \citenamefont
  {Zelan}, \citenamefont {Hill~III}, \citenamefont {Lobb}, \citenamefont
  {Helmerson}, \citenamefont {Phillips},\ and\ \citenamefont
  {Campbell}}]{Ramanathan2011}%
  \BibitemOpen
  \bibfield  {author} {\bibinfo {author} {\bibfnamefont {A.}~\bibnamefont
  {Ramanathan}}, \bibinfo {author} {\bibfnamefont {K.~C.}\ \bibnamefont
  {Wright}}, \bibinfo {author} {\bibfnamefont {S.~R.}\ \bibnamefont {Muniz}},
  \bibinfo {author} {\bibfnamefont {M.}~\bibnamefont {Zelan}}, \bibinfo
  {author} {\bibfnamefont {W.~T.}\ \bibnamefont {Hill~III}}, \bibinfo {author}
  {\bibfnamefont {C.~J.}\ \bibnamefont {Lobb}}, \bibinfo {author}
  {\bibfnamefont {K.}~\bibnamefont {Helmerson}}, \bibinfo {author}
  {\bibfnamefont {W.~D.}\ \bibnamefont {Phillips}}, \ and\ \bibinfo {author}
  {\bibfnamefont {G.~K.}\ \bibnamefont {Campbell}},\ }\bibfield  {title}
  {\bibinfo {title} {\emph {Superflow in a toroidal Bose-Einstein condensate:
  an atom circuit with a tunable weak link}},\ }\href@noop {} {\bibfield
  {journal} {\bibinfo  {journal} {Phys. Rev. Lett.}\ }\textbf {\bibinfo
  {volume} {106}},\ \bibinfo {pages} {130401} (\bibinfo {year}
  {2011})}\BibitemShut {NoStop}%
\bibitem [{\citenamefont {Ryu}\ \emph {et~al.}(2013)\citenamefont {Ryu},
  \citenamefont {Blackburn}, \citenamefont {Blinova},\ and\ \citenamefont
  {Boshier}}]{Ryu2013}%
  \BibitemOpen
  \bibfield  {author} {\bibinfo {author} {\bibfnamefont {C.}~\bibnamefont
  {Ryu}}, \bibinfo {author} {\bibfnamefont {P.~W.}\ \bibnamefont {Blackburn}},
  \bibinfo {author} {\bibfnamefont {A.~A.}\ \bibnamefont {Blinova}}, \ and\
  \bibinfo {author} {\bibfnamefont {M.~G.}\ \bibnamefont {Boshier}},\
  }\bibfield  {title} {\bibinfo {title} {\emph {Experimental realization of
  Josephson junctions for an atom SQUID}},\ }\href@noop {} {\bibfield
  {journal} {\bibinfo  {journal} {Phys. Rev. Lett.}\ }\textbf {\bibinfo
  {volume} {111}},\ \bibinfo {pages} {205301} (\bibinfo {year}
  {2013})}\BibitemShut {NoStop}%
\bibitem [{\citenamefont {Eckel}\ \emph
  {et~al.}(2014{\natexlab{a}})\citenamefont {Eckel}, \citenamefont {Lee},
  \citenamefont {Jendrzejewski}, \citenamefont {Murray}, \citenamefont {Clark},
  \citenamefont {Lobb}, \citenamefont {Phillips}, \citenamefont {Edwards},\
  and\ \citenamefont {Campbell}}]{Eckel2014_hysteresis}%
  \BibitemOpen
  \bibfield  {author} {\bibinfo {author} {\bibfnamefont {S.}~\bibnamefont
  {Eckel}}, \bibinfo {author} {\bibfnamefont {J.~G.}\ \bibnamefont {Lee}},
  \bibinfo {author} {\bibfnamefont {F.}~\bibnamefont {Jendrzejewski}}, \bibinfo
  {author} {\bibfnamefont {N.}~\bibnamefont {Murray}}, \bibinfo {author}
  {\bibfnamefont {C.~W.}\ \bibnamefont {Clark}}, \bibinfo {author}
  {\bibfnamefont {C.~J.}\ \bibnamefont {Lobb}}, \bibinfo {author}
  {\bibfnamefont {W.~D.}\ \bibnamefont {Phillips}}, \bibinfo {author}
  {\bibfnamefont {M.}~\bibnamefont {Edwards}}, \ and\ \bibinfo {author}
  {\bibfnamefont {G.~K.}\ \bibnamefont {Campbell}},\ }\bibfield  {title}
  {\bibinfo {title} {\emph {Hysteresis in a quantized superfluid
  ‘atomtronic’circuit}},\ }\href@noop {} {\bibfield  {journal} {\bibinfo
  {journal} {Nature}\ }\textbf {\bibinfo {volume} {506}},\ \bibinfo {pages}
  {200} (\bibinfo {year} {2014}{\natexlab{a}})}\BibitemShut {NoStop}%
\bibitem [{\citenamefont {Ryu}\ \emph {et~al.}(2020)\citenamefont {Ryu},
  \citenamefont {Samson},\ and\ \citenamefont {Boshier}}]{Ryu2020}%
  \BibitemOpen
  \bibfield  {author} {\bibinfo {author} {\bibfnamefont {C.}~\bibnamefont
  {Ryu}}, \bibinfo {author} {\bibfnamefont {E.}~\bibnamefont {Samson}}, \ and\
  \bibinfo {author} {\bibfnamefont {M.~G.}\ \bibnamefont {Boshier}},\
  }\bibfield  {title} {\bibinfo {title} {\emph {Quantum interference of
  currents in an atomtronic SQUID}},\ }\href@noop {} {\bibfield  {journal}
  {\bibinfo  {journal} {Nature communications}\ }\textbf {\bibinfo {volume}
  {11}},\ \bibinfo {pages} {1} (\bibinfo {year} {2020})}\BibitemShut {NoStop}%
\bibitem [{\citenamefont {Kiehn}\ \emph {et~al.}(2022)\citenamefont {Kiehn},
  \citenamefont {Singh},\ and\ \citenamefont {Mathey}}]{Kiehn2022}%
  \BibitemOpen
  \bibfield  {author} {\bibinfo {author} {\bibfnamefont {H.}~\bibnamefont
  {Kiehn}}, \bibinfo {author} {\bibfnamefont {V.~P.}\ \bibnamefont {Singh}}, \
  and\ \bibinfo {author} {\bibfnamefont {L.}~\bibnamefont {Mathey}},\
  }\bibfield  {title} {\bibinfo {title} {\emph {Implementation of an atomtronic
  SQUID in a strongly confined toroidal condensate}},\ }\href {\doibase
  10.1103/PhysRevResearch.4.033024} {\bibfield  {journal} {\bibinfo  {journal}
  {Phys. Rev. Research}\ }\textbf {\bibinfo {volume} {4}},\ \bibinfo {pages}
  {033024} (\bibinfo {year} {2022})}\BibitemShut {NoStop}%
\bibitem [{\citenamefont {Gustavson}\ \emph {et~al.}(1997)\citenamefont
  {Gustavson}, \citenamefont {Bouyer},\ and\ \citenamefont
  {Kasevich}}]{Gustavson1997}%
  \BibitemOpen
  \bibfield  {author} {\bibinfo {author} {\bibfnamefont {T.~L.}\ \bibnamefont
  {Gustavson}}, \bibinfo {author} {\bibfnamefont {P.}~\bibnamefont {Bouyer}}, \
  and\ \bibinfo {author} {\bibfnamefont {M.~A.}\ \bibnamefont {Kasevich}},\
  }\bibfield  {title} {\bibinfo {title} {\emph {Precision rotation measurements
  with an atom interferometer gyroscope}},\ }\href@noop {} {\bibfield
  {journal} {\bibinfo  {journal} {Phys. Rev. Lett.}\ }\textbf {\bibinfo
  {volume} {78}},\ \bibinfo {pages} {2046} (\bibinfo {year}
  {1997})}\BibitemShut {NoStop}%
\bibitem [{\citenamefont {Navez}\ \emph {et~al.}(2016)\citenamefont {Navez},
  \citenamefont {Pandey}, \citenamefont {Mas}, \citenamefont {Poulios},
  \citenamefont {Fernholz},\ and\ \citenamefont {von Klitzing}}]{Navez2016}%
  \BibitemOpen
  \bibfield  {author} {\bibinfo {author} {\bibfnamefont {P.}~\bibnamefont
  {Navez}}, \bibinfo {author} {\bibfnamefont {S.}~\bibnamefont {Pandey}},
  \bibinfo {author} {\bibfnamefont {H.}~\bibnamefont {Mas}}, \bibinfo {author}
  {\bibfnamefont {K.}~\bibnamefont {Poulios}}, \bibinfo {author} {\bibfnamefont
  {T.}~\bibnamefont {Fernholz}}, \ and\ \bibinfo {author} {\bibfnamefont
  {W.}~\bibnamefont {von Klitzing}},\ }\bibfield  {title} {\bibinfo {title}
  {\emph {Matter-wave interferometers using TAAP rings}},\ }\href@noop {}
  {\bibfield  {journal} {\bibinfo  {journal} {New J. Phys.}\ }\textbf {\bibinfo
  {volume} {18}},\ \bibinfo {pages} {075014} (\bibinfo {year}
  {2016})}\BibitemShut {NoStop}%
\bibitem [{\citenamefont {Pandey}\ \emph {et~al.}(2019)\citenamefont {Pandey},
  \citenamefont {Mas}, \citenamefont {Drougakis}, \citenamefont {Thekkeppatt},
  \citenamefont {Bolpasi}, \citenamefont {Vasilakis}, \citenamefont {Poulios},\
  and\ \citenamefont {von Klitzing}}]{Pandey2019}%
  \BibitemOpen
  \bibfield  {author} {\bibinfo {author} {\bibfnamefont {S.}~\bibnamefont
  {Pandey}}, \bibinfo {author} {\bibfnamefont {H.}~\bibnamefont {Mas}},
  \bibinfo {author} {\bibfnamefont {G.}~\bibnamefont {Drougakis}}, \bibinfo
  {author} {\bibfnamefont {P.}~\bibnamefont {Thekkeppatt}}, \bibinfo {author}
  {\bibfnamefont {V.}~\bibnamefont {Bolpasi}}, \bibinfo {author} {\bibfnamefont
  {G.}~\bibnamefont {Vasilakis}}, \bibinfo {author} {\bibfnamefont
  {K.}~\bibnamefont {Poulios}}, \ and\ \bibinfo {author} {\bibfnamefont
  {W.}~\bibnamefont {von Klitzing}},\ }\bibfield  {title} {\bibinfo {title}
  {\emph {Hypersonic Bose--Einstein condensates in accelerator rings}},\ }\href
  {\doibase 10.1038/s41586-019-1273-5} {\bibfield  {journal} {\bibinfo
  {journal} {Nature}\ }\textbf {\bibinfo {volume} {570}},\ \bibinfo {pages}
  {205} (\bibinfo {year} {2019})}\BibitemShut {NoStop}%
\bibitem [{\citenamefont {Hofferberth}\ \emph {et~al.}(2007)\citenamefont
  {Hofferberth}, \citenamefont {Lesanovsky}, \citenamefont {Fischer},
  \citenamefont {Schumm},\ and\ \citenamefont
  {Schmiedmayer}}]{Hofferberth2007}%
  \BibitemOpen
  \bibfield  {author} {\bibinfo {author} {\bibfnamefont {S.}~\bibnamefont
  {Hofferberth}}, \bibinfo {author} {\bibfnamefont {I.}~\bibnamefont
  {Lesanovsky}}, \bibinfo {author} {\bibfnamefont {B.}~\bibnamefont {Fischer}},
  \bibinfo {author} {\bibfnamefont {T.}~\bibnamefont {Schumm}}, \ and\ \bibinfo
  {author} {\bibfnamefont {J.}~\bibnamefont {Schmiedmayer}},\ }\bibfield
  {title} {\bibinfo {title} {\emph {Non-equilibrium coherence dynamics in
  one-dimensional Bose gases}},\ }\href@noop {} {\bibfield  {journal} {\bibinfo
   {journal} {Nature}\ }\textbf {\bibinfo {volume} {449}},\ \bibinfo {pages}
  {324} (\bibinfo {year} {2007})}\BibitemShut {NoStop}%
\bibitem [{\citenamefont {Schweigler}\ \emph {et~al.}(2017)\citenamefont
  {Schweigler}, \citenamefont {Kasper}, \citenamefont {Erne}, \citenamefont
  {Mazets}, \citenamefont {Rauer}, \citenamefont {Cataldini}, \citenamefont
  {Langen}, \citenamefont {Gasenzer}, \citenamefont {Berges},\ and\
  \citenamefont {Schmiedmayer}}]{Schweigler2017}%
  \BibitemOpen
  \bibfield  {author} {\bibinfo {author} {\bibfnamefont {T.}~\bibnamefont
  {Schweigler}}, \bibinfo {author} {\bibfnamefont {V.}~\bibnamefont {Kasper}},
  \bibinfo {author} {\bibfnamefont {S.}~\bibnamefont {Erne}}, \bibinfo {author}
  {\bibfnamefont {I.}~\bibnamefont {Mazets}}, \bibinfo {author} {\bibfnamefont
  {B.}~\bibnamefont {Rauer}}, \bibinfo {author} {\bibfnamefont
  {F.}~\bibnamefont {Cataldini}}, \bibinfo {author} {\bibfnamefont
  {T.}~\bibnamefont {Langen}}, \bibinfo {author} {\bibfnamefont
  {T.}~\bibnamefont {Gasenzer}}, \bibinfo {author} {\bibfnamefont
  {J.}~\bibnamefont {Berges}}, \ and\ \bibinfo {author} {\bibfnamefont
  {J.}~\bibnamefont {Schmiedmayer}},\ }\bibfield  {title} {\bibinfo {title}
  {\emph {Experimental characterization of a quantum many-body system via
  higher-order correlations}},\ }\href {\doibase 10.1038/nature22310}
  {\bibfield  {journal} {\bibinfo  {journal} {Nature}\ }\textbf {\bibinfo
  {volume} {545}},\ \bibinfo {pages} {323} (\bibinfo {year}
  {2017})}\BibitemShut {NoStop}%
\bibitem [{\citenamefont {Burger}\ \emph {et~al.}(1999)\citenamefont {Burger},
  \citenamefont {Bongs}, \citenamefont {Dettmer}, \citenamefont {Ertmer},
  \citenamefont {Sengstock}, \citenamefont {Sanpera}, \citenamefont
  {Shlyapnikov},\ and\ \citenamefont {Lewenstein}}]{Burger1999}%
  \BibitemOpen
  \bibfield  {author} {\bibinfo {author} {\bibfnamefont {S.}~\bibnamefont
  {Burger}}, \bibinfo {author} {\bibfnamefont {K.}~\bibnamefont {Bongs}},
  \bibinfo {author} {\bibfnamefont {S.}~\bibnamefont {Dettmer}}, \bibinfo
  {author} {\bibfnamefont {W.}~\bibnamefont {Ertmer}}, \bibinfo {author}
  {\bibfnamefont {K.}~\bibnamefont {Sengstock}}, \bibinfo {author}
  {\bibfnamefont {A.}~\bibnamefont {Sanpera}}, \bibinfo {author} {\bibfnamefont
  {G.~V.}\ \bibnamefont {Shlyapnikov}}, \ and\ \bibinfo {author} {\bibfnamefont
  {M.}~\bibnamefont {Lewenstein}},\ }\bibfield  {title} {\bibinfo {title}
  {\emph {Dark solitons in Bose-Einstein condensates}},\ }\href@noop {}
  {\bibfield  {journal} {\bibinfo  {journal} {Physical Review Letters}\
  }\textbf {\bibinfo {volume} {83}},\ \bibinfo {pages} {5198} (\bibinfo {year}
  {1999})}\BibitemShut {NoStop}%
\bibitem [{\citenamefont {Yefsah}\ \emph {et~al.}(2013)\citenamefont {Yefsah},
  \citenamefont {Sommer}, \citenamefont {Ku}, \citenamefont {Cheuk},
  \citenamefont {Ji}, \citenamefont {Bakr},\ and\ \citenamefont
  {Zwierlein}}]{Yefsah2013}%
  \BibitemOpen
  \bibfield  {author} {\bibinfo {author} {\bibfnamefont {T.}~\bibnamefont
  {Yefsah}}, \bibinfo {author} {\bibfnamefont {A.~T.}\ \bibnamefont {Sommer}},
  \bibinfo {author} {\bibfnamefont {M.~J.}\ \bibnamefont {Ku}}, \bibinfo
  {author} {\bibfnamefont {L.~W.}\ \bibnamefont {Cheuk}}, \bibinfo {author}
  {\bibfnamefont {W.}~\bibnamefont {Ji}}, \bibinfo {author} {\bibfnamefont
  {W.~S.}\ \bibnamefont {Bakr}}, \ and\ \bibinfo {author} {\bibfnamefont
  {M.~W.}\ \bibnamefont {Zwierlein}},\ }\bibfield  {title} {\bibinfo {title}
  {\emph {Heavy solitons in a fermionic superfluid}},\ }\href@noop {}
  {\bibfield  {journal} {\bibinfo  {journal} {Nature}\ }\textbf {\bibinfo
  {volume} {499}},\ \bibinfo {pages} {426} (\bibinfo {year}
  {2013})}\BibitemShut {NoStop}%
\bibitem [{\citenamefont {Luick}\ \emph {et~al.}(2020)\citenamefont {Luick},
  \citenamefont {Sobirey}, \citenamefont {Bohlen}, \citenamefont {Singh},
  \citenamefont {Mathey}, \citenamefont {Lompe},\ and\ \citenamefont
  {Moritz}}]{Luick2020}%
  \BibitemOpen
  \bibfield  {author} {\bibinfo {author} {\bibfnamefont {N.}~\bibnamefont
  {Luick}}, \bibinfo {author} {\bibfnamefont {L.}~\bibnamefont {Sobirey}},
  \bibinfo {author} {\bibfnamefont {M.}~\bibnamefont {Bohlen}}, \bibinfo
  {author} {\bibfnamefont {V.~P.}\ \bibnamefont {Singh}}, \bibinfo {author}
  {\bibfnamefont {L.}~\bibnamefont {Mathey}}, \bibinfo {author} {\bibfnamefont
  {T.}~\bibnamefont {Lompe}}, \ and\ \bibinfo {author} {\bibfnamefont
  {H.}~\bibnamefont {Moritz}},\ }\bibfield  {title} {\bibinfo {title} {\emph
  {An ideal Josephson junction in an ultracold two-dimensional Fermi gas}},\
  }\href@noop {} {\bibfield  {journal} {\bibinfo  {journal} {Science}\ }\textbf
  {\bibinfo {volume} {369}},\ \bibinfo {pages} {89} (\bibinfo {year}
  {2020})}\BibitemShut {NoStop}%
\bibitem [{\citenamefont {Zheng}\ and\ \citenamefont
  {Javanainen}(2003)}]{Zheng2003}%
  \BibitemOpen
  \bibfield  {author} {\bibinfo {author} {\bibfnamefont {Y.}~\bibnamefont
  {Zheng}}\ and\ \bibinfo {author} {\bibfnamefont {J.}~\bibnamefont
  {Javanainen}},\ }\bibfield  {title} {\bibinfo {title} {\emph {Classical and
  quantum models for phase imprinting}},\ }\href@noop {} {\bibfield  {journal}
  {\bibinfo  {journal} {Phys. Rev. A}\ }\textbf {\bibinfo {volume} {67}},\
  \bibinfo {pages} {035602} (\bibinfo {year} {2003})}\BibitemShut {NoStop}%
\bibitem [{\citenamefont {Eckel}\ \emph
  {et~al.}(2014{\natexlab{b}})\citenamefont {Eckel}, \citenamefont
  {Jendrzejewski}, \citenamefont {Kumar}, \citenamefont {Lobb},\ and\
  \citenamefont {Campbell}}]{Eckel_2014}%
  \BibitemOpen
  \bibfield  {author} {\bibinfo {author} {\bibfnamefont {S.}~\bibnamefont
  {Eckel}}, \bibinfo {author} {\bibfnamefont {F.}~\bibnamefont
  {Jendrzejewski}}, \bibinfo {author} {\bibfnamefont {A.}~\bibnamefont
  {Kumar}}, \bibinfo {author} {\bibfnamefont {C.~J.}\ \bibnamefont {Lobb}}, \
  and\ \bibinfo {author} {\bibfnamefont {G.~K.}\ \bibnamefont {Campbell}},\
  }\bibfield  {title} {\bibinfo {title} {\emph {Interferometric measurement of
  the current-phase relationship of a superfluid weak link}},\ }\href@noop {}
  {\bibfield  {journal} {\bibinfo  {journal} {Phys. Rev. X}\ }\textbf {\bibinfo
  {volume} {4}},\ \bibinfo {pages} {031052} (\bibinfo {year}
  {2014}{\natexlab{b}})}\BibitemShut {NoStop}%
\bibitem [{\citenamefont {Corman}\ \emph {et~al.}(2014)\citenamefont {Corman},
  \citenamefont {Chomaz}, \citenamefont {Bienaim{\'e}}, \citenamefont
  {Desbuquois}, \citenamefont {Weitenberg}, \citenamefont {Nascimbene},
  \citenamefont {Dalibard},\ and\ \citenamefont {Beugnon}}]{Corman2014}%
  \BibitemOpen
  \bibfield  {author} {\bibinfo {author} {\bibfnamefont {L.}~\bibnamefont
  {Corman}}, \bibinfo {author} {\bibfnamefont {L.}~\bibnamefont {Chomaz}},
  \bibinfo {author} {\bibfnamefont {T.}~\bibnamefont {Bienaim{\'e}}}, \bibinfo
  {author} {\bibfnamefont {R.}~\bibnamefont {Desbuquois}}, \bibinfo {author}
  {\bibfnamefont {C.}~\bibnamefont {Weitenberg}}, \bibinfo {author}
  {\bibfnamefont {S.}~\bibnamefont {Nascimbene}}, \bibinfo {author}
  {\bibfnamefont {J.}~\bibnamefont {Dalibard}}, \ and\ \bibinfo {author}
  {\bibfnamefont {J.}~\bibnamefont {Beugnon}},\ }\bibfield  {title} {\bibinfo
  {title} {\emph {Quench-induced supercurrents in an annular Bose gas}},\
  }\href@noop {} {\bibfield  {journal} {\bibinfo  {journal} {Phys. Rev. Lett.}\
  }\textbf {\bibinfo {volume} {113}},\ \bibinfo {pages} {135302} (\bibinfo
  {year} {2014})}\BibitemShut {NoStop}%
\bibitem [{\citenamefont {Mathew}\ \emph {et~al.}(2015)\citenamefont {Mathew},
  \citenamefont {Kumar}, \citenamefont {Eckel}, \citenamefont {Jendrzejewski},
  \citenamefont {Campbell}, \citenamefont {Edwards},\ and\ \citenamefont
  {Tiesinga}}]{Mathew2015}%
  \BibitemOpen
  \bibfield  {author} {\bibinfo {author} {\bibfnamefont {R.}~\bibnamefont
  {Mathew}}, \bibinfo {author} {\bibfnamefont {A.}~\bibnamefont {Kumar}},
  \bibinfo {author} {\bibfnamefont {S.}~\bibnamefont {Eckel}}, \bibinfo
  {author} {\bibfnamefont {F.}~\bibnamefont {Jendrzejewski}}, \bibinfo {author}
  {\bibfnamefont {G.~K.}\ \bibnamefont {Campbell}}, \bibinfo {author}
  {\bibfnamefont {M.}~\bibnamefont {Edwards}}, \ and\ \bibinfo {author}
  {\bibfnamefont {E.}~\bibnamefont {Tiesinga}},\ }\bibfield  {title} {\bibinfo
  {title} {\emph {Self-heterodyne detection of the in situ phase of an atomic
  superconducting quantum interference device}},\ }\href@noop {} {\bibfield
  {journal} {\bibinfo  {journal} {Phys. Rev. A}\ }\textbf {\bibinfo {volume}
  {92}},\ \bibinfo {pages} {033602} (\bibinfo {year} {2015})}\BibitemShut
  {NoStop}%
\bibitem [{SM()}]{SM}%
  \BibitemOpen
  \href@noop {} {}\bibinfo {note} {See Supplemental Material for further
  details on the experimental methods and the theoretical modeling
  tools.}\BibitemShut {Stop}%
\bibitem [{\citenamefont {P\'erez-Obiol}\ \emph {et~al.}(2022)\citenamefont
  {P\'erez-Obiol}, \citenamefont {Polo},\ and\ \citenamefont
  {Amico}}]{Perez2021}%
  \BibitemOpen
  \bibfield  {author} {\bibinfo {author} {\bibfnamefont {A.}~\bibnamefont
  {P\'erez-Obiol}}, \bibinfo {author} {\bibfnamefont {J.}~\bibnamefont {Polo}},
  \ and\ \bibinfo {author} {\bibfnamefont {L.}~\bibnamefont {Amico}},\
  }\bibfield  {title} {\bibinfo {title} {\emph {Coherent phase slips in coupled
  matter-wave circuits}},\ }\href {\doibase 10.1103/PhysRevResearch.4.L022038}
  {\bibfield  {journal} {\bibinfo  {journal} {Phys. Rev. Research}\ }\textbf
  {\bibinfo {volume} {4}},\ \bibinfo {pages} {L022038} (\bibinfo {year}
  {2022})}\BibitemShut {NoStop}%
\bibitem [{\citenamefont {Bland}\ \emph {et~al.}(2020)\citenamefont {Bland},
  \citenamefont {Marolleau}, \citenamefont {Comaron}, \citenamefont {Malomed},\
  and\ \citenamefont {Proukakis}}]{Bland2020}%
  \BibitemOpen
  \bibfield  {author} {\bibinfo {author} {\bibfnamefont {T.}~\bibnamefont
  {Bland}}, \bibinfo {author} {\bibfnamefont {Q.}~\bibnamefont {Marolleau}},
  \bibinfo {author} {\bibfnamefont {P.}~\bibnamefont {Comaron}}, \bibinfo
  {author} {\bibfnamefont {B.}~\bibnamefont {Malomed}}, \ and\ \bibinfo
  {author} {\bibfnamefont {N.}~\bibnamefont {Proukakis}},\ }\bibfield  {title}
  {\bibinfo {title} {\emph {Persistent current formation in double-ring
  geometries}},\ }\href@noop {} {\bibfield  {journal} {\bibinfo  {journal} {J.
  Phys. B}\ }\textbf {\bibinfo {volume} {53}},\ \bibinfo {pages} {115301}
  (\bibinfo {year} {2020})}\BibitemShut {NoStop}%
\bibitem [{\citenamefont {Pelegr{\'\i}}\ \emph {et~al.}(2019)\citenamefont
  {Pelegr{\'\i}}, \citenamefont {Marques}, \citenamefont {Ahufinger},
  \citenamefont {Mompart},\ and\ \citenamefont {Dias}}]{Pelegri2019}%
  \BibitemOpen
  \bibfield  {author} {\bibinfo {author} {\bibfnamefont {G.}~\bibnamefont
  {Pelegr{\'\i}}}, \bibinfo {author} {\bibfnamefont {A.~M.}\ \bibnamefont
  {Marques}}, \bibinfo {author} {\bibfnamefont {V.}~\bibnamefont {Ahufinger}},
  \bibinfo {author} {\bibfnamefont {J.}~\bibnamefont {Mompart}}, \ and\
  \bibinfo {author} {\bibfnamefont {R.~G.}\ \bibnamefont {Dias}},\ }\bibfield
  {title} {\bibinfo {title} {\emph {Second-order topological corner states with
  ultracold atoms carrying orbital angular momentum in optical lattices}},\
  }\href@noop {} {\bibfield  {journal} {\bibinfo  {journal} {Phys. Rev. B}\
  }\textbf {\bibinfo {volume} {100}},\ \bibinfo {pages} {205109} (\bibinfo
  {year} {2019})}\BibitemShut {NoStop}%
\bibitem [{\citenamefont {Pecci}\ \emph {et~al.}(2021)\citenamefont {Pecci},
  \citenamefont {Naldesi}, \citenamefont {Amico},\ and\ \citenamefont
  {Minguzzi}}]{Pecci2021}%
  \BibitemOpen
  \bibfield  {author} {\bibinfo {author} {\bibfnamefont {G.}~\bibnamefont
  {Pecci}}, \bibinfo {author} {\bibfnamefont {P.}~\bibnamefont {Naldesi}},
  \bibinfo {author} {\bibfnamefont {L.}~\bibnamefont {Amico}}, \ and\ \bibinfo
  {author} {\bibfnamefont {A.}~\bibnamefont {Minguzzi}},\ }\bibfield  {title}
  {\bibinfo {title} {\emph {Probing the BCS-BEC crossover with persistent
  currents}},\ }\href@noop {} {\bibfield  {journal} {\bibinfo  {journal} {Phys.
  Rev. Research}\ }\textbf {\bibinfo {volume} {3}},\ \bibinfo {pages} {L032064}
  (\bibinfo {year} {2021})}\BibitemShut {NoStop}%
\bibitem [{\citenamefont {Sacha}\ and\ \citenamefont
  {Delande}(2014)}]{Sacha2014}%
  \BibitemOpen
  \bibfield  {author} {\bibinfo {author} {\bibfnamefont {K.}~\bibnamefont
  {Sacha}}\ and\ \bibinfo {author} {\bibfnamefont {D.}~\bibnamefont
  {Delande}},\ }\bibfield  {title} {\bibinfo {title} {\emph {Proper phase
  imprinting method for a dark soliton excitation in a superfluid Fermi
  mixture}},\ }\href@noop {} {\bibfield  {journal} {\bibinfo  {journal}
  {Physical Review A}\ }\textbf {\bibinfo {volume} {90}},\ \bibinfo {pages}
  {021604(R)} (\bibinfo {year} {2014})}\BibitemShut {NoStop}%
\bibitem [{\citenamefont {Kumar}\ \emph {et~al.}(2018)\citenamefont {Kumar},
  \citenamefont {Dubessy}, \citenamefont {Badr}, \citenamefont {De~Rossi},
  \citenamefont {de~Go\"er~de Herve}, \citenamefont {Longchambon},\ and\
  \citenamefont {Perrin}}]{Perrin2018}%
  \BibitemOpen
  \bibfield  {author} {\bibinfo {author} {\bibfnamefont {A.}~\bibnamefont
  {Kumar}}, \bibinfo {author} {\bibfnamefont {R.}~\bibnamefont {Dubessy}},
  \bibinfo {author} {\bibfnamefont {T.}~\bibnamefont {Badr}}, \bibinfo {author}
  {\bibfnamefont {C.}~\bibnamefont {De~Rossi}}, \bibinfo {author}
  {\bibfnamefont {M.}~\bibnamefont {de~Go\"er~de Herve}}, \bibinfo {author}
  {\bibfnamefont {L.}~\bibnamefont {Longchambon}}, \ and\ \bibinfo {author}
  {\bibfnamefont {H.}~\bibnamefont {Perrin}},\ }\bibfield  {title} {\bibinfo
  {title} {\emph {Producing superfluid circulation states using phase
  imprinting}},\ }\href {\doibase 10.1103/PhysRevA.97.043615} {\bibfield
  {journal} {\bibinfo  {journal} {Phys. Rev. A}\ }\textbf {\bibinfo {volume}
  {97}},\ \bibinfo {pages} {043615} (\bibinfo {year} {2018})}\BibitemShut
  {NoStop}%
\bibitem [{\citenamefont {Xhani}\ \emph {et~al.}(2020)\citenamefont {Xhani},
  \citenamefont {Neri}, \citenamefont {Galantucci}, \citenamefont {Scazza},
  \citenamefont {Burchianti}, \citenamefont {Lee}, \citenamefont {Barenghi},
  \citenamefont {Trombettoni}, \citenamefont {Inguscio}, \citenamefont
  {Zaccanti}, \citenamefont {Roati},\ and\ \citenamefont
  {Proukakis}}]{Xhani2020}%
  \BibitemOpen
  \bibfield  {author} {\bibinfo {author} {\bibfnamefont {K.}~\bibnamefont
  {Xhani}}, \bibinfo {author} {\bibfnamefont {E.}~\bibnamefont {Neri}},
  \bibinfo {author} {\bibfnamefont {L.}~\bibnamefont {Galantucci}}, \bibinfo
  {author} {\bibfnamefont {F.}~\bibnamefont {Scazza}}, \bibinfo {author}
  {\bibfnamefont {A.}~\bibnamefont {Burchianti}}, \bibinfo {author}
  {\bibfnamefont {K.-L.}\ \bibnamefont {Lee}}, \bibinfo {author} {\bibfnamefont
  {C.~F.}\ \bibnamefont {Barenghi}}, \bibinfo {author} {\bibfnamefont
  {A.}~\bibnamefont {Trombettoni}}, \bibinfo {author} {\bibfnamefont
  {M.}~\bibnamefont {Inguscio}}, \bibinfo {author} {\bibfnamefont
  {M.}~\bibnamefont {Zaccanti}}, \bibinfo {author} {\bibfnamefont
  {G.}~\bibnamefont {Roati}}, \ and\ \bibinfo {author} {\bibfnamefont {N.~P.}\
  \bibnamefont {Proukakis}},\ }\bibfield  {title} {\bibinfo {title} {\emph
  {Critical Transport and Vortex Dynamics in a Thin Atomic Josephson
  Junction}},\ }\href {\doibase 10.1103/PhysRevLett.124.045301} {\bibfield
  {journal} {\bibinfo  {journal} {Phys. Rev. Lett.}\ }\textbf {\bibinfo
  {volume} {124}},\ \bibinfo {pages} {045301} (\bibinfo {year}
  {2020})}\BibitemShut {NoStop}%
\bibitem [{\citenamefont {Reeves}\ \emph {et~al.}(2015)\citenamefont {Reeves},
  \citenamefont {Billam}, \citenamefont {Anderson},\ and\ \citenamefont
  {Bradley}}]{Reeves2015}%
  \BibitemOpen
  \bibfield  {author} {\bibinfo {author} {\bibfnamefont {M.~T.}\ \bibnamefont
  {Reeves}}, \bibinfo {author} {\bibfnamefont {T.~P.}\ \bibnamefont {Billam}},
  \bibinfo {author} {\bibfnamefont {B.~P.}\ \bibnamefont {Anderson}}, \ and\
  \bibinfo {author} {\bibfnamefont {A.~S.}\ \bibnamefont {Bradley}},\
  }\bibfield  {title} {\bibinfo {title} {\emph {Identifying a superfluid
  Reynolds number via dynamical similarity}},\ }\href@noop {} {\bibfield
  {journal} {\bibinfo  {journal} {Phys. Rev. Lett.}\ }\textbf {\bibinfo
  {volume} {114}},\ \bibinfo {pages} {155302} (\bibinfo {year}
  {2015})}\BibitemShut {NoStop}%
\bibitem [{\citenamefont {Kwon}\ \emph {et~al.}(2015)\citenamefont {Kwon},
  \citenamefont {Seo},\ and\ \citenamefont {Shin}}]{Kwon2015}%
  \BibitemOpen
  \bibfield  {author} {\bibinfo {author} {\bibfnamefont {W.~J.}\ \bibnamefont
  {Kwon}}, \bibinfo {author} {\bibfnamefont {S.~W.}\ \bibnamefont {Seo}}, \
  and\ \bibinfo {author} {\bibfnamefont {Y.-i.}\ \bibnamefont {Shin}},\
  }\bibfield  {title} {\bibinfo {title} {\emph {Periodic shedding of vortex
  dipoles from a moving penetrable obstacle in a Bose-Einstein condensate}},\
  }\href@noop {} {\bibfield  {journal} {\bibinfo  {journal} {Phys. Rev. A}\
  }\textbf {\bibinfo {volume} {92}},\ \bibinfo {pages} {033613} (\bibinfo
  {year} {2015})}\BibitemShut {NoStop}%
\bibitem [{\citenamefont {Park}\ \emph {et~al.}(2018)\citenamefont {Park},
  \citenamefont {Ko},\ and\ \citenamefont {Shin}}]{Park2018}%
  \BibitemOpen
  \bibfield  {author} {\bibinfo {author} {\bibfnamefont {J.~W.}\ \bibnamefont
  {Park}}, \bibinfo {author} {\bibfnamefont {B.}~\bibnamefont {Ko}}, \ and\
  \bibinfo {author} {\bibfnamefont {Y.-i.}\ \bibnamefont {Shin}},\ }\bibfield
  {title} {\bibinfo {title} {\emph {Critical vortex shedding in a strongly
  interacting fermionic superfluid}},\ }\href@noop {} {\bibfield  {journal}
  {\bibinfo  {journal} {Phys. Rev. Lett.}\ }\textbf {\bibinfo {volume} {121}},\
  \bibinfo {pages} {225301} (\bibinfo {year} {2018})}\BibitemShut {NoStop}%
\bibitem [{\citenamefont {Giamarchi}(2016)}]{Giamarchi2016}%
  \BibitemOpen
  \bibfield  {author} {\bibinfo {author} {\bibfnamefont {T.}~\bibnamefont
  {Giamarchi}},\ }\bibfield  {title} {\bibinfo {title} {\emph {Current drag in
  two leg quantum ladders}},\ }\href {\doibase
  https://doi.org/10.1016/j.physe.2016.01.019} {\bibfield  {journal} {\bibinfo
  {journal} {Phys. E: Low-Dimens. Syst. Nanostructures}\ }\textbf {\bibinfo
  {volume} {82}},\ \bibinfo {pages} {66} (\bibinfo {year} {2016})},\ \bibinfo
  {note} {{Frontiers in quantum electronic transport - In memory of Markus
  B\"uttiker}}\BibitemShut {NoStop}%
\bibitem [{\citenamefont {Chestnov}\ \emph {et~al.}(2021)\citenamefont
  {Chestnov}, \citenamefont {Yulin}, \citenamefont {Shelykh},\ and\
  \citenamefont {Kavokin}}]{Chestnov2021}%
  \BibitemOpen
  \bibfield  {author} {\bibinfo {author} {\bibfnamefont {I.}~\bibnamefont
  {Chestnov}}, \bibinfo {author} {\bibfnamefont {A.}~\bibnamefont {Yulin}},
  \bibinfo {author} {\bibfnamefont {I.~A.}\ \bibnamefont {Shelykh}}, \ and\
  \bibinfo {author} {\bibfnamefont {A.}~\bibnamefont {Kavokin}},\ }\bibfield
  {title} {\bibinfo {title} {\emph {Dissipative Josephson vortices in annular
  polariton fluids}},\ }\href {\doibase 10.1103/PhysRevB.104.165305} {\bibfield
   {journal} {\bibinfo  {journal} {Phys. Rev. B}\ }\textbf {\bibinfo {volume}
  {104}},\ \bibinfo {pages} {165305} (\bibinfo {year} {2021})}\BibitemShut
  {NoStop}%
\bibitem [{\citenamefont {Mooij}\ and\ \citenamefont
  {Harmans}(2005)}]{Mooij2005}%
  \BibitemOpen
  \bibfield  {author} {\bibinfo {author} {\bibfnamefont {J.}~\bibnamefont
  {Mooij}}\ and\ \bibinfo {author} {\bibfnamefont {C.}~\bibnamefont
  {Harmans}},\ }\bibfield  {title} {\bibinfo {title} {\emph {Phase-slip flux
  qubits}},\ }\href@noop {} {\bibfield  {journal} {\bibinfo  {journal} {New J.
  Phys.}\ }\textbf {\bibinfo {volume} {7}},\ \bibinfo {pages} {219} (\bibinfo
  {year} {2005})}\BibitemShut {NoStop}%
\bibitem [{\citenamefont {Ragole}\ and\ \citenamefont
  {Taylor}(2016)}]{Ragole2016}%
  \BibitemOpen
  \bibfield  {author} {\bibinfo {author} {\bibfnamefont {S.}~\bibnamefont
  {Ragole}}\ and\ \bibinfo {author} {\bibfnamefont {J.~M.}\ \bibnamefont
  {Taylor}},\ }\bibfield  {title} {\bibinfo {title} {\emph {Interacting atomic
  interferometry for rotation sensing approaching the Heisenberg Limit}},\
  }\href@noop {} {\bibfield  {journal} {\bibinfo  {journal} {Phys. Rev. Lett.}\
  }\textbf {\bibinfo {volume} {117}},\ \bibinfo {pages} {203002} (\bibinfo
  {year} {2016})}\BibitemShut {NoStop}%
\bibitem [{\citenamefont {Kwon}\ \emph {et~al.}(2021)\citenamefont {Kwon},
  \citenamefont {Del~Pace}, \citenamefont {Xhani}, \citenamefont {Galantucci},
  \citenamefont {Falconi}, \citenamefont {Inguscio}, \citenamefont {Scazza},\
  and\ \citenamefont {Roati}}]{Kwon2021}%
  \BibitemOpen
  \bibfield  {author} {\bibinfo {author} {\bibfnamefont {W.}~\bibnamefont
  {Kwon}}, \bibinfo {author} {\bibfnamefont {G.}~\bibnamefont {Del~Pace}},
  \bibinfo {author} {\bibfnamefont {K.}~\bibnamefont {Xhani}}, \bibinfo
  {author} {\bibfnamefont {L.}~\bibnamefont {Galantucci}}, \bibinfo {author}
  {\bibfnamefont {A.~M.}\ \bibnamefont {Falconi}}, \bibinfo {author}
  {\bibfnamefont {M.}~\bibnamefont {Inguscio}}, \bibinfo {author}
  {\bibfnamefont {F.}~\bibnamefont {Scazza}}, \ and\ \bibinfo {author}
  {\bibfnamefont {G.}~\bibnamefont {Roati}},\ }\bibfield  {title} {\bibinfo
  {title} {\emph {Sound emission and annihilations in a programmable quantum
  vortex collider}},\ }\href@noop {} {\bibfield  {journal} {\bibinfo  {journal}
  {Nature}\ }\textbf {\bibinfo {volume} {6}},\ \bibinfo {pages} {64} (\bibinfo
  {year} {2021})}\BibitemShut {NoStop}%
\bibitem [{\citenamefont {Combescot}\ \emph {et~al.}(2006)\citenamefont
  {Combescot}, \citenamefont {Kagan},\ and\ \citenamefont
  {Stringari}}]{Comberscot2006}%
  \BibitemOpen
  \bibfield  {author} {\bibinfo {author} {\bibfnamefont {R.}~\bibnamefont
  {Combescot}}, \bibinfo {author} {\bibfnamefont {M.~Y.}\ \bibnamefont
  {Kagan}}, \ and\ \bibinfo {author} {\bibfnamefont {S.}~\bibnamefont
  {Stringari}},\ }\bibfield  {title} {\bibinfo {title} {\emph {Collective mode
  of homogeneous superfluid Fermi gases in the BEC-BCS crossover}},\ }\href
  {\doibase 10.1103/PhysRevA.74.042717} {\bibfield  {journal} {\bibinfo
  {journal} {Phys. Rev. A}\ }\textbf {\bibinfo {volume} {74}},\ \bibinfo
  {pages} {042717} (\bibinfo {year} {2006})}\BibitemShut {NoStop}%
\bibitem [{\citenamefont {Weimer}\ \emph {et~al.}(2015)\citenamefont {Weimer},
  \citenamefont {Morgener}, \citenamefont {Singh}, \citenamefont {Siegl},
  \citenamefont {Hueck}, \citenamefont {Luick}, \citenamefont {Mathey},\ and\
  \citenamefont {Moritz}}]{Weimer2015}%
  \BibitemOpen
  \bibfield  {author} {\bibinfo {author} {\bibfnamefont {W.}~\bibnamefont
  {Weimer}}, \bibinfo {author} {\bibfnamefont {K.}~\bibnamefont {Morgener}},
  \bibinfo {author} {\bibfnamefont {V.~P.}\ \bibnamefont {Singh}}, \bibinfo
  {author} {\bibfnamefont {J.}~\bibnamefont {Siegl}}, \bibinfo {author}
  {\bibfnamefont {K.}~\bibnamefont {Hueck}}, \bibinfo {author} {\bibfnamefont
  {N.}~\bibnamefont {Luick}}, \bibinfo {author} {\bibfnamefont
  {L.}~\bibnamefont {Mathey}}, \ and\ \bibinfo {author} {\bibfnamefont
  {H.}~\bibnamefont {Moritz}},\ }\bibfield  {title} {\bibinfo {title} {\emph
  {Critical velocity in the BEC-BCS crossover}},\ }\href@noop {} {\bibfield
  {journal} {\bibinfo  {journal} {Phys. Rev. Lett.}\ }\textbf {\bibinfo
  {volume} {114}},\ \bibinfo {pages} {095301} (\bibinfo {year}
  {2015})}\BibitemShut {NoStop}%
\bibitem [{\citenamefont {Ketterle}\ and\ \citenamefont
  {Zwierlein}(2008)}]{ketterle2008}%
  \BibitemOpen
  \bibfield  {author} {\bibinfo {author} {\bibfnamefont {W.}~\bibnamefont
  {Ketterle}}\ and\ \bibinfo {author} {\bibfnamefont {M.~W.}\ \bibnamefont
  {Zwierlein}},\ }\bibfield  {title} {\bibinfo {title} {\emph {Making, probing
  and understanding ultracold Fermi gases}},\ }\href@noop {} {\bibfield
  {journal} {\bibinfo  {journal} {La Rivista del Nuovo Cimento}\ }\textbf
  {\bibinfo {volume} {31}},\ \bibinfo {pages} {247} (\bibinfo {year}
  {2008})}\BibitemShut {NoStop}%
\bibitem [{\citenamefont {Kwon}\ \emph {et~al.}(2020)\citenamefont {Kwon},
  \citenamefont {Del~Pace}, \citenamefont {Panza}, \citenamefont {Inguscio},
  \citenamefont {Zwerger}, \citenamefont {Zaccanti}, \citenamefont {Scazza},\
  and\ \citenamefont {Roati}}]{Kwon2020}%
  \BibitemOpen
  \bibfield  {author} {\bibinfo {author} {\bibfnamefont {W.~J.}\ \bibnamefont
  {Kwon}}, \bibinfo {author} {\bibfnamefont {G.}~\bibnamefont {Del~Pace}},
  \bibinfo {author} {\bibfnamefont {R.}~\bibnamefont {Panza}}, \bibinfo
  {author} {\bibfnamefont {M.}~\bibnamefont {Inguscio}}, \bibinfo {author}
  {\bibfnamefont {W.}~\bibnamefont {Zwerger}}, \bibinfo {author} {\bibfnamefont
  {M.}~\bibnamefont {Zaccanti}}, \bibinfo {author} {\bibfnamefont
  {F.}~\bibnamefont {Scazza}}, \ and\ \bibinfo {author} {\bibfnamefont
  {G.}~\bibnamefont {Roati}},\ }\bibfield  {title} {\bibinfo {title} {\emph
  {Strongly correlated superfluid order parameters from dc Josephson
  supercurrents}},\ }\href {\doibase 10.1126/science.aaz2463} {\bibfield
  {journal} {\bibinfo  {journal} {Science}\ }\textbf {\bibinfo {volume}
  {369}},\ \bibinfo {pages} {84} (\bibinfo {year} {2020})}\BibitemShut
  {NoStop}%
\bibitem [{\citenamefont {Hoskinson}\ \emph {et~al.}(2006)\citenamefont
  {Hoskinson}, \citenamefont {Sato}, \citenamefont {Hahn},\ and\ \citenamefont
  {Packard}}]{hoskinson-2006}%
  \BibitemOpen
  \bibfield  {author} {\bibinfo {author} {\bibfnamefont {E.}~\bibnamefont
  {Hoskinson}}, \bibinfo {author} {\bibfnamefont {Y.}~\bibnamefont {Sato}},
  \bibinfo {author} {\bibfnamefont {I.}~\bibnamefont {Hahn}}, \ and\ \bibinfo
  {author} {\bibfnamefont {R.~E.}\ \bibnamefont {Packard}},\ }\bibfield
  {title} {\bibinfo {title} {\emph {Transition from phase slips to the
  Josephson effect in a superfluid ${}^4$He weak link}},\ }\href
  {https://doi.org/10.1038/nphys19} {\bibfield  {journal} {\bibinfo  {journal}
  {Nat. Phys.}\ }\textbf {\bibinfo {volume} {2}},\ \bibinfo {pages} {23}
  (\bibinfo {year} {2006})}\BibitemShut {NoStop}%
\bibitem [{\citenamefont {Sato}\ and\ \citenamefont
  {Packard}(2011)}]{sato-packard-2012}%
  \BibitemOpen
  \bibfield  {author} {\bibinfo {author} {\bibfnamefont {Y.}~\bibnamefont
  {Sato}}\ and\ \bibinfo {author} {\bibfnamefont {R.~E.}\ \bibnamefont
  {Packard}},\ }\bibfield  {title} {\bibinfo {title} {\emph {Superfluid helium
  quantum interference devices: physics and applications}},\ }\href {\doibase
  10.1088/0034-4885/75/1/016401} {\bibfield  {journal} {\bibinfo  {journal}
  {Rep. Prog. Phys.}\ }\textbf {\bibinfo {volume} {75}},\ \bibinfo {pages}
  {016401} (\bibinfo {year} {2011})}\BibitemShut {NoStop}%
\bibitem [{\citenamefont {Donadello}\ \emph {et~al.}(2014)\citenamefont
  {Donadello}, \citenamefont {Serafini}, \citenamefont {Tylutki}, \citenamefont
  {Pitaevskii}, \citenamefont {Dalfovo}, \citenamefont {Lamporesi},\ and\
  \citenamefont {Ferrari}}]{Donadello2014}%
  \BibitemOpen
  \bibfield  {author} {\bibinfo {author} {\bibfnamefont {S.}~\bibnamefont
  {Donadello}}, \bibinfo {author} {\bibfnamefont {S.}~\bibnamefont {Serafini}},
  \bibinfo {author} {\bibfnamefont {M.}~\bibnamefont {Tylutki}}, \bibinfo
  {author} {\bibfnamefont {L.~P.}\ \bibnamefont {Pitaevskii}}, \bibinfo
  {author} {\bibfnamefont {F.}~\bibnamefont {Dalfovo}}, \bibinfo {author}
  {\bibfnamefont {G.}~\bibnamefont {Lamporesi}}, \ and\ \bibinfo {author}
  {\bibfnamefont {G.}~\bibnamefont {Ferrari}},\ }\bibfield  {title} {\bibinfo
  {title} {\emph {Observation of solitonic vortices in Bose-Einstein
  condensates}},\ }\href@noop {} {\bibfield  {journal} {\bibinfo  {journal}
  {Physical review letters}\ }\textbf {\bibinfo {volume} {113}},\ \bibinfo
  {pages} {065302} (\bibinfo {year} {2014})}\BibitemShut {NoStop}%
\bibitem [{\citenamefont {Ku}\ \emph {et~al.}(2016)\citenamefont {Ku},
  \citenamefont {Mukherjee}, \citenamefont {Yefsah},\ and\ \citenamefont
  {Zwierlein}}]{ku2016cascade}%
  \BibitemOpen
  \bibfield  {author} {\bibinfo {author} {\bibfnamefont {M.~J.~H.}\
  \bibnamefont {Ku}}, \bibinfo {author} {\bibfnamefont {B.}~\bibnamefont
  {Mukherjee}}, \bibinfo {author} {\bibfnamefont {T.}~\bibnamefont {Yefsah}}, \
  and\ \bibinfo {author} {\bibfnamefont {M.~W.}\ \bibnamefont {Zwierlein}},\
  }\bibfield  {title} {\bibinfo {title} {\emph {Cascade of solitonic
  excitations in a superfluid fermi gas: From planar solitons to vortex rings
  and lines}},\ }\href@noop {} {\bibfield  {journal} {\bibinfo  {journal}
  {Physical review letters}\ }\textbf {\bibinfo {volume} {116}},\ \bibinfo
  {pages} {045304} (\bibinfo {year} {2016})}\BibitemShut {NoStop}%
\end{thebibliography}

\begin{thebibliography}{6}%
\makeatletter
\providecommand \@ifxundefined [1]{%
 \@ifx{#1\undefined}
}%
\providecommand \@ifnum [1]{%
 \ifnum #1\expandafter \@firstoftwo
 \else \expandafter \@secondoftwo
 \fi
}%
\providecommand \@ifx [1]{%
 \ifx #1\expandafter \@firstoftwo
 \else \expandafter \@secondoftwo
 \fi
}%
\providecommand \natexlab [1]{#1}%
\providecommand \enquote  [1]{``#1''}%
\providecommand \bibnamefont  [1]{#1}%
\providecommand \bibfnamefont [1]{#1}%
\providecommand \citenamefont [1]{#1}%
\providecommand \href@noop [0]{\@secondoftwo}%
\providecommand \href [0]{\begingroup \@sanitize@url \@href}%
\providecommand \@href[1]{\@@startlink{#1}\@@href}%
\providecommand \@@href[1]{\endgroup#1\@@endlink}%
\providecommand \@sanitize@url [0]{\catcode `\\12\catcode `\$12\catcode
  `\&12\catcode `\#12\catcode `\^12\catcode `\_12\catcode `\%12\relax}%
\providecommand \@@startlink[1]{}%
\providecommand \@@endlink[0]{}%
\providecommand \url  [0]{\begingroup\@sanitize@url \@url }%
\providecommand \@url [1]{\endgroup\@href {#1}{\urlprefix }}%
\providecommand \urlprefix  [0]{URL }%
\providecommand \Eprint [0]{\href }%
\providecommand \doibase [0]{http://dx.doi.org/}%
\providecommand \selectlanguage [0]{\@gobble}%
\providecommand \bibinfo  [0]{\@secondoftwo}%
\providecommand \bibfield  [0]{\@secondoftwo}%
\providecommand \translation [1]{[#1]}%
\providecommand \BibitemOpen [0]{}%
\providecommand \bibitemStop [0]{}%
\providecommand \bibitemNoStop [0]{.\EOS\space}%
\providecommand \EOS [0]{\spacefactor3000\relax}%
\providecommand \BibitemShut  [1]{\csname bibitem#1\endcsname}%
\let\auto@bib@innerbib\@empty
\bibitem [{\citenamefont {Kwon}\ \emph {et~al.}(2020)\citenamefont {Kwon},
  \citenamefont {Del~Pace}, \citenamefont {Panza}, \citenamefont {Inguscio},
  \citenamefont {Zwerger}, \citenamefont {Zaccanti}, \citenamefont {Scazza},\
  and\ \citenamefont {Roati}}]{Kwon2020}%
  \BibitemOpen
  \bibfield  {author} {\bibinfo {author} {\bibfnamefont {W.~J.}\ \bibnamefont
  {Kwon}}, \bibinfo {author} {\bibfnamefont {G.}~\bibnamefont {Del~Pace}},
  \bibinfo {author} {\bibfnamefont {R.}~\bibnamefont {Panza}}, \bibinfo
  {author} {\bibfnamefont {M.}~\bibnamefont {Inguscio}}, \bibinfo {author}
  {\bibfnamefont {W.}~\bibnamefont {Zwerger}}, \bibinfo {author} {\bibfnamefont
  {M.}~\bibnamefont {Zaccanti}}, \bibinfo {author} {\bibfnamefont
  {F.}~\bibnamefont {Scazza}}, \ and\ \bibinfo {author} {\bibfnamefont
  {G.}~\bibnamefont {Roati}},\ }\bibfield  {title} {\bibinfo {title} {\emph
  {Strongly correlated superfluid order parameters from dc Josephson
  supercurrents}},\ }\href {\doibase 10.1126/science.aaz2463} {\bibfield
  {journal} {\bibinfo  {journal} {Science}\ }\textbf {\bibinfo {volume}
  {369}},\ \bibinfo {pages} {84} (\bibinfo {year} {2020})}\BibitemShut
  {NoStop}%
\bibitem [{\citenamefont {Griffin}\ \emph {et~al.}(2009)\citenamefont
  {Griffin}, \citenamefont {Nikuni},\ and\ \citenamefont {Zaremba}}]{ZNG}%
  \BibitemOpen
  \bibfield  {author} {\bibinfo {author} {\bibfnamefont {A.}~\bibnamefont
  {Griffin}}, \bibinfo {author} {\bibfnamefont {T.}~\bibnamefont {Nikuni}}, \
  and\ \bibinfo {author} {\bibfnamefont {E.}~\bibnamefont {Zaremba}},\ }\href
  {\doibase 10.1017/CBO9780511575150} {\emph {\bibinfo {title} {Bose-Condensed
  Gases at Finite Temperatures}}}\ (\bibinfo  {publisher} {Cambridge University
  Press},\ \bibinfo {year} {2009})\BibitemShut {NoStop}%
\bibitem [{\citenamefont {Xhani}\ and\ \citenamefont
  {Proukakis}(2022)}]{Xhani2022}%
  \BibitemOpen
  \bibfield  {author} {\bibinfo {author} {\bibfnamefont {K.}~\bibnamefont
  {Xhani}}\ and\ \bibinfo {author} {\bibfnamefont {N.~P.}\ \bibnamefont
  {Proukakis}},\ }\bibfield  {title} {\bibinfo {title} {\emph {Dissipation in a
  finite-temperature atomic Josephson junction}},\ }\href {\doibase
  10.1103/PhysRevResearch.4.033205} {\bibfield  {journal} {\bibinfo  {journal}
  {Phys. Rev. Research}\ }\textbf {\bibinfo {volume} {4}},\ \bibinfo {pages}
  {033205} (\bibinfo {year} {2022})}\BibitemShut {NoStop}%
\bibitem [{\citenamefont {Heiselberg}(2004)}]{Heiselberg2004}%
  \BibitemOpen
  \bibfield  {author} {\bibinfo {author} {\bibfnamefont {H.}~\bibnamefont
  {Heiselberg}},\ }\bibfield  {title} {\bibinfo {title} {\emph {Collective
  Modes of Trapped Gases at the BEC-BCS Crossover}},\ }\href {\doibase
  10.1103/PhysRevLett.93.040402} {\bibfield  {journal} {\bibinfo  {journal}
  {Phys. Rev. Lett.}\ }\textbf {\bibinfo {volume} {93}},\ \bibinfo {pages}
  {040402} (\bibinfo {year} {2004})}\BibitemShut {NoStop}%
\bibitem [{\citenamefont {Haussmann}\ and\ \citenamefont
  {Zwerger}(2008)}]{Haussmann2008}%
  \BibitemOpen
  \bibfield  {author} {\bibinfo {author} {\bibfnamefont {R.}~\bibnamefont
  {Haussmann}}\ and\ \bibinfo {author} {\bibfnamefont {W.}~\bibnamefont
  {Zwerger}},\ }\bibfield  {title} {\bibinfo {title} {\emph {Thermodynamics of
  a trapped unitary Fermi gas}},\ }\href {\doibase 10.1103/PhysRevA.78.063602}
  {\bibfield  {journal} {\bibinfo  {journal} {Phys. Rev. A}\ }\textbf {\bibinfo
  {volume} {78}},\ \bibinfo {pages} {063602} (\bibinfo {year}
  {2008})}\BibitemShut {NoStop}%
\bibitem [{\citenamefont {Haussmann}\ \emph {et~al.}(2007)\citenamefont
  {Haussmann}, \citenamefont {Rantner}, \citenamefont {Cerrito},\ and\
  \citenamefont {Zwerger}}]{Haussmann2007}%
  \BibitemOpen
  \bibfield  {author} {\bibinfo {author} {\bibfnamefont {R.}~\bibnamefont
  {Haussmann}}, \bibinfo {author} {\bibfnamefont {W.}~\bibnamefont {Rantner}},
  \bibinfo {author} {\bibfnamefont {S.}~\bibnamefont {Cerrito}}, \ and\
  \bibinfo {author} {\bibfnamefont {W.}~\bibnamefont {Zwerger}},\ }\bibfield
  {title} {\bibinfo {title} {\emph {Thermodynamics of the BCS-BEC crossover}},\
  }\href@noop {} {\bibfield  {journal} {\bibinfo  {journal} {Phys. Rev. A}\
  }\textbf {\bibinfo {volume} {75}},\ \bibinfo {pages} {023610} (\bibinfo
  {year} {2007})}\BibitemShut {NoStop}%
\end{thebibliography}
\end{document}